\documentclass[english]{iopart}
				   
\begin{document}

\title[Galois group for the Ising model]
{\Large
Square lattice Ising model susceptibility:
connection matrices and singular behavior of $\chi^{(3)}$ and $\chi^{(4)}$}
 
\author{ 
N. Zenine$^\S$, S. Boukraa$^\dag$, S. Hassani$^\S$ and
J.-M. Maillard$^\ddag$}
\address{\S  Centre de Recherche Nucl\'eaire d'Alger, \\
2 Bd. Frantz Fanon, BP 399, 16000 Alger, Algeria}
\address{\dag Universit\'e de Blida, Institut d'A{\'e}ronautique,
 Blida, Algeria}
\address{\ddag\ LPTMC, Universit\'e de Paris 6, Tour 24,
 4\`eme \'etage, case 121, \\
 4 Place Jussieu, 75252 Paris Cedex 05, France} 
\ead{maillard@lptmc.jussieu.fr, maillard@lptl.jussieu.fr, 
sboukraa@wissal.dz, njzenine@yahoo.com}

\begin{abstract}
We present a simple, but efficient, way to calculate
connection matrices between sets of independent local solutions, defined at
two neighboring singular points, of Fuchsian differential equations of
quite large orders, such as those found for the
 third and fourth contribution ($\chi^{(3)}$ and $\chi^{(4)}$)
 to the magnetic susceptibility of the square lattice Ising model.
We deduce all the critical behavior of the solutions 
$\chi^{(3)}$ and $\chi^{(4)}$, as well as the asymptotic behavior 
of the coefficients in the corresponding series expansions.
We confirm that the newly found quadratic singularities of
the  Fuchsian ODE associated with $\chi^{(3)}$ are not singularities of
the  particular solution $\chi^{(3)}$ itself.
 We use the previous connection matrices
to get the exact expressions of all the monodromy matrices 
of the Fuchsian differential equation for $\chi^{(3)}$ (and $\chi^{(4)}$)
expressed in the same basis of solutions.
These monodromy  matrices  are the generators of the differential Galois
 group of the Fuchsian
differential equations for $\chi^{(3)}$ (and $\chi^{(4)}$),
whose analysis is just sketched here.
As far as the physics implications of the solutions are concerned, we 
find challenging qualitative differences when comparing the corrections
to scaling
for the full susceptibillity $\chi$ at high temperature (resp. low temperature)
and the first two terms $\chi^{(1)}$ and $\chi^{(3)}$
 (resp. $\chi^{(2)}$ and $\chi^{(4)}$) .
\end{abstract} 
\vskip .5cm


\noindent {\bf AMS Classification scheme numbers}: 34M55, 
47E05, 81Qxx, 32G34, 34Lxx, 34Mxx, 14Kxx


\section{Introduction}
\label{intro}

Since a pioneering, and quite monumental, paper~\cite{wu-mc-tr-ba-76}
on the two-dimensional  Ising models, it has been known that 
the magnetic susceptibility of square lattice Ising model,
can be written \cite{wu-mc-tr-ba-76} as an infinite sum
of $(n-1)$-dimensional 
integrals~\cite{nappi-78,pal-tra-81,yamada-84,yamada-85,nickel-99,nickel-00}
 contributions:
\begin{eqnarray}
\chi (T)\,=\,\,\sum_{n=1}^{\infty }\chi^{(n)}(T) 
\end{eqnarray}
 The odd (respectively even) $n$ correspond to the 
high (respectively low) temperature domain.
These $(n-1)$-dimensional integrals
are known to be holonomic, since they are integrals
 of holonomic (actually algebraic)
integrands. Besides the known $\, \chi^{(1)}$ and
$\, \chi^{(2)}$ terms, which can be expressed in
 terms of simple algebraic or hypergeometric
functions, it is only recently that the Fuchsian differential equations
satisfied  by the $\, \chi^{(3)}$ and
$\, \chi^{(4)}$ terms have been 
found~\cite{ze-bo-ha-ma-04,ze-bo-ha-ma-05,ze-bo-ha-ma-05b}.
These two exact differential equations of quite large orders (seven and ten)
can be used to find answers 
to a set of problems traditionally known to be subtle, and difficult, for 
functions with confluent singularities, like the fine-tuning of
 the singular behaviors
 for all the singularities (dominant singular behavior, 
sub-dominant, etc.), accurate calculations  
of the asymptotic behavior of the coefficients, etc.

Recall that the third, and fourth, contribution to the magnetic
susceptibility $\chi^{(3)}$, and $\chi^{(4)}$, are given by multi-integrals
and each is, thus, a particular solution of the corresponding differential
equation. These differential equations exhibit a finite set of regular
singular points that may (or may not) appear in the physical solutions
$\chi^{(3)}$ and $\chi^{(4)}$. Besides the physical singularities
and the non physical singularities
$s=\pm i$ (where $\, s=\sinh(2K)$, $K$ being  the usual Ising model
 coupling constant, $\, K\, = \, \beta \, J$),
it is commonly believed
that the $\chi^{(n)}$'s have, at least, other non physical singularities
given by B. Nickel \cite{nickel-99, nickel-00}.
The dominant singular behaviors at all these (non physical) singularities
($\chi^{(3)}$ and $\chi^{(4)}$) have also been given by B. Nickel.
 The differential equations of the $\chi^{(n)}$'s,
which ``encode'' all the information on the solutions and their singular
behavior, in fact, allow us to obtain not only the dominant, but also
all the subdominant singular behavior, hardly detectable from straight series
analysis.
It is thus of interest to get (or confirm) these
singular behaviors from the exact Fuchsian differential equations
that we have actually obtained for
$\chi^{(3)}$ and $\chi^{(4)}$ and, especially, the singular behavior
at the two new quadratic singularities, $\, 1\, +3\, w\, + 4\, w^ 2\, = \, 0$,
(where $\, w=s/(1+s^2)/2$) 
found  for $\chi^{(3)}$ \cite{ze-bo-ha-ma-04}.

The physical solution $\chi^{(3)}$ is defined by a double integral on
two angles and is known as a series
 obtained by expansion (then integration) of the double
integral at $w=0$ (or $s=0$).
It is certainly not simple to obtain the $\chi^{(3)}$ expansion around (say)
the ferromagnetic critical point $w=1/4$, due to a singular logarithmic
behavior.
However, one can overcome this difficulty since, 
with a differential equation, it is straightforward to obtain the formal
series solutions at each regular singular point (i.e., a local basis of series solutions).
By connecting the formal solutions
around $w=0$ and the formal series solutions around another regular singular point
like $w=1/4$, one will be able to express the particular solution $\chi^{(3)}$ (and also
all the other formal solutions) as a linear combination of solutions
valid at $w=1/4$.
The seven local solutions at $w=0$ will, then,  be given by the product of
a $7 \times 7$ matrix with the vector having the seven local
solutions at $w=1/4$ as entries.
In other words, succeeding in obtaining these connection matrices
amounts to building a common (global) basis 
of solutions valid for all the regular singular points.
Furthermore, with these connection matrices,
we obtain, in fact, the analytic continuation in the whole complex plane
of the variable $w$, of $\chi^{(3)}$ and $\chi^{(4)}$,
which are known as integral representations.

Note that, remarkably, the Fuchsian differential equation for $\chi^{(3)}$
  has simple rational, and algebraic, solutions.
These rational or algebraic solutions, known in closed form, 
can be understood  {\em globally}.
One can easily expand such  {\em globally defined} solutions
around any singular point of the ODE, and follow 
these solutions through any ``jump'' from one regular singularity
to another one, and, therefore, from one
well-suited basis to another well-suited basis.
For a function not known in closed form, like the ``physical'' solution
$\, \chi^{(3)}$, the decomposition 
on each well-suited local basis associated with every 
singular point of the ODE, is far from clear.
The correspondence between these various (well-suited) local bases 
associated with each singular point of the ODE,
is typically a  {\em global} problem and, thus, a quite difficult one.
One clearly needs to build
{\em effective methods} to find such connection matrices 
in the case of Fuchsian differential equations
of order seven, or ten ($\chi^{(3)}$ and $\, \chi^{(4)}$), 
or of much higher orders ($\chi^{(5)}$,
$\, \chi^{(6)}$, etc.).
With a method of matching of series, we will show that
the connection matrices matching these
various well-suited bases of series-solutions
can be obtained explicitly. 
The entries of these matrices can be calculated with
{\em as many digits as we want}.
We will show that we can actually find the exact expressions of 
these entries as simple algebraic expressions
of (in the case of the Fuchsian ODE's of $\, \chi^{(3)}$ and $\, \chi^{(4)}$)
powers of $\, \pi$, $\ln(2)$, $\ln(3)$ and various algebraic numbers 
or integers, together with more ``transcendental" numbers like
the "ferromagnetic constant" $\, I_3^{+}$
introduced in equation (7.12) of~\cite{wu-mc-tr-ba-76}:
\begin{eqnarray} 
 I_3^{+} &=& {{1} \over {2\, \pi^2}} \cdot 
\int_{1}^{ \infty}\int_{1}^{\infty} \int_{1}^{ \infty} 
dy_1\, dy_2\, dy_3 \cdot \Bigl({{y_2^2\, -1 }
 \over {(y_1^2\, -1)\,(y_3^2\, -1)  }}\Bigr)^{1/2} \cdot 
Y^2       \nonumber \\
\label{I3p}
& =&  0.0008144625656625044393912171285627219978\, \cdots
 \\
&& Y \, = \, {{ y_1\, -y_3} \over { 
(y_1\, +y_2)\,(y_2\, +y_3)\,(y_1\, +y_2\, +y_3)  }}\nonumber
\end{eqnarray} 
Focusing on $\chi^{(3)}$, and since this physical solution is known as
a series expansion at $w=0$ (low or high temperature expansions), 
we will give all the connection matrices between this  $w=0$ regular singular point
and all the other regular singularities of the differential equation
including the two new complex regular
singularities~\cite{ze-bo-ha-ma-04,ze-bo-ha-ma-05} which are
 roots of $\, 1+3w+4w^2=0$.
We will comment on the occurrence of
the "ferromagnetic constant" $\, I_3^{+}$ 
in the various blocks of the connection matrices.
The decomposition of $\, \chi^{(3)}$ in the well-suited basis for each
regular singular point allows us to find all the singular behavior
of the physical solution.
 From these results, we will deduce the
asymptotic behavior of the coefficients of the series expansion  of
 $\, \chi^{(3)}$.
 These last problems
are interesting, per se, for series expansions analysis of lattice 
statistical mechanics, since 
they correspond to subtle analysis of confluent singularities.
Actually, we
will see that even the 
last asymptotic evaluation problem is a (global) connection problem
since the physical solution  like $\, \chi^{(3)}$ does not correspond to the
 obvious dominant singular behavior 
one might have imagined from the indicial equation.

Focusing on the two new singularities, the roots of $\, 1+3w+4w^2=0$, we will show
that the physical solution $\chi^{(3)}$ {\em is not singular at these points}.
The factor of the logarithmic term,
in the decomposition of $\chi^{(3)}$ at these singular points,
is known exactly and vanishes identically.

Note that a fundamental concept to understand
 (the symmetries, the solutions of) these
exact Fuchsian differential equations is the so-called {\em differential
Galois group}~\cite{Moscou}.
Differential  Galois groups have been calculated
for simple enough second order, or even third order, ODE's (see
 for instance~\cite{SingUlm}). However, finding 
the differential Galois group of such higher order
Fuchsian differential equations (order seven for $\chi^{(3)}$, order ten for $\chi^{(4)}$)
 with eight regular singular points (for $\chi^{(3)}$)
is not an easy task~\cite{SingUlm} and
requires the computation of all the monodromy matrices
associated with each (non apparent) regular singular point,
considered {\em in the same basis}\footnote[2]{These monodromy matrices
are the generators of the monodromy group which identifies with the
differential Galois group when there are no irregular singularities,
and, thus, no Stokes matrices~\cite{Stokes}.}.

We will give the exact expression of all the monodromy matrices
{\em expressed in the same} ($w=0$) {\em basis of solutions},
these eight matrices being the generators of the differential
Galois group,
which will be given in a
forthcoming publication~\cite{Weil}.

This method can be generalized,
{\em mutatis mutandis}, to the Fuchsian differential equation of $\, \chi^{(4)}$.
Here, we give the connection matrix between $w=0$ and,
both, the ferromagnetic, and anti-ferromagnetic, critical points.
The singular behavior is straightforwardly obtained with the asymptotic
behavior of the series coefficients of the physical solution $\chi^{(4)}$.
The monodromy matrices, expressed in the {\em same basis} of solutions
are also obtained.

The paper is organized as follows. We recall, in Section 2,
some results on the Fuchsian differential
equation satisfied by $\, \chi^{(3)}$, and give a {\em new factorization} for the corresponding
order seven differential operator, yielding the emergence
of an order two, and an order three, differential operator
(denoted $\, Z_2$ and $\, Y_3$ below).
We give, in Section 3, the connection matrices matching the
(series) solutions around the regular singular point $\, w=0$
and around all the other regular singular points.
With these connection matrices we deduce the  singularity
behavior and the asymptotics on the physical solution of
this ODE (Section 4).
In Section 5, we  deduce the exact expressions
of the monodromy matrices expressed {\em in the same basis}.
Section 6 generalizes these results to the Fuchsian differential
equation satisfied by $\, \chi^{(4)}$.
Some physics implications of our results at scaling are
discussed in Section 7.
Our conclusion is given in Section 8.

\section{The order seven operator $L_7$}
\label{L7again}
Let us first recall, with the same notations as
 in~\cite{ze-bo-ha-ma-04,ze-bo-ha-ma-05},
the seven linearly independent solutions given in
\cite{ze-bo-ha-ma-04,ze-bo-ha-ma-05} for the order seven differential
operator  $L_7$, associated with\footnote[3]{$\tilde{\chi}^{(n)}$ is
defined as $\chi^{(n)}=(1-s^4)^{1/4}/s \cdot \tilde{\chi}^{(n)}$,
for $n$ odd.}
$\, \tilde{\chi}^{(3)}$.

One finds two  remarkable {\em rational, and algebraic,}
solutions of the order seven differential equation associated with
$\, \tilde{\chi}^{(3)}$, namely:
\begin{eqnarray}
\label{remarkable}
{\cal S}(L_1) \, = \, \, {{w} \over {1\, -4\, w}}, \qquad \qquad 
{\cal S}(N_1) \, = \, \,
{\frac {{w}^{2}}{ \left( 1-4\,w \right) \sqrt {1-16\,{w}^{2}}}} \qquad 
\end{eqnarray}
associated with the two order $1$ differential operators given in~\cite{ze-bo-ha-ma-04}:
\begin{eqnarray}
\label{L1N1}
L_1 \, = \, \, {{d} \over {dw}} \,  -{{1} \over {w\, (1-4\, w)}}, \qquad 
N_1 \, = \, \, {{d} \over {dw}} \, -{{2\, (1+2\, w)} \over {w\, (1-16\, w^2)}}
\end{eqnarray}
There is a solution  behaving like $\, w^3$, that we denote $\, S_3$:
\begin{eqnarray}
\label{S3}
&& S_3 \, = \, \, 
{w}^{3}+3\,{w}^{4}+22\,{w}^{5}+74\,{w}^{6}+417\,{w}^{7}+1465\,{w}^{8}
\nonumber \\
&& \qquad \qquad +7479\,{w}^{9}  +26839\,{w}^{10}\, + \cdots 
\end{eqnarray}
and three solutions
with logarithmic terms given by equation (17) in \cite{ze-bo-ha-ma-04}.
Note the singled-out series expansion starting with $\, w^9$,
corresponding to the physical solution $\, \tilde{\chi}^{(3)}$:
\begin{eqnarray}
\label{singledout}
S_9 \, = \, \, {{\tilde{\chi}^{(3)} (w)} \over {8}}  \, = \, \, 
{w}^{9}+36\,{w}^{11}
+4\,{w}^{12}+884\,{w}^{13}+196\,{w}^{14}\, + \cdots
\end{eqnarray}

The choice of this set of linearly independent solutions (and of
these series) is, in fact, arbitrary since any
linear combination of solutions is also a solution of the differential equation.
Three of the above solutions are however singled out: the solutions
${\cal S}(L_1)$ and ${\cal S}(N_1)$ which are {\em global} (since they have closed expression),
 and the series $S_9$ associated with the
highest critical exponent in the indicial equation ($w^ 9\, + \cdots $), 
which has a unique (well-defined) expression 
and happens to
correspond to the ``physical'' solution $\tilde{\chi}^{(3)}$. Linear combinations, 
like $S_3\, -\alpha \cdot S_9$, are, at first sight, on the same footing.

Nevertheless, introducing such a specific linear combination, 
B. Nickel\footnote[4]{We thank B. Nickel for kindly communicating this
result.} has been able to show that the
resulting series for the particular value
$\alpha=16$ is, also, the solution
of a linear differential equation of lower order, namely order four.
With this result, the factorization scheme of $L_7$
becomes\footnote[2]{The order four differential operator found by B. Nickel
corresponds to
$B_2 \cdot T_1 \cdot  L_1=\, B_2 \cdot O_1 \cdot N_1=X_1 \cdot Z_2 \cdot N_1$.} :
\begin{eqnarray}
\label{morefactoL7}
L_7 \, &=& \, M_1 \cdot Y_3 \cdot Z_2 \cdot N_1   
\, = \, B_3 \cdot X_1 \cdot Z_2 \cdot N_1   \\
\, &=& \, B_3 \cdot B_2 \cdot O_1 \cdot N_1
\, = \, B_3 \cdot B_2 \cdot T_1 \cdot L_1  \nonumber
\end{eqnarray}
where the indices correspond to the order of 
the differential operators ($B_3, \, Y_3\, $ are order three, $B_2, \, Z_2\, $  order two, ...).
The differential operators $L_7$,  $M_1$ and $T_1$
have been given in \cite{ze-bo-ha-ma-04}.
We give in Appendix A,
the differential operators $X_1$, $Z_2$ and  $Y_3$.
With these differential operators, all the factorizations (\ref{morefactoL7})
can be found by left and right division.

From these factorizations of $\,L_7$, one can see that the general solution
of the corresponding differential equation is the {\em direct} sum of the
solution of $L_1$ and of the general solution of the differential operator
$L_6= Y_3 \cdot Z_2 \cdot N_1$. The operator $\, L_{7}$ has the following
 decomposition:
\begin{eqnarray}
\label{factochi3}
 L_{7} \,  = \, \,  L_{6} \oplus  L_1. 
\end{eqnarray}
We thus consider, from now on, the differential operator $L_6$.

The formal solutions of $L_6$ (at the singular point $w=0$) show
the occurrence of
three Frobenius series and three solutions carrying logarithmic terms.
With the factorizations (\ref{morefactoL7}),
it is interesting to see which operator brings with it
a singular behavior for a given regular singular point.
Table 1 shows the critical exponents at each regular singular
point for both differential operators $Z_2 \cdot N_1$ and
$\, Y_3 \cdot Z_2 \cdot N_1$.
In the third and sixth column
the number of independent solutions with logarithmic terms is shown.

At the singular points $w=1$, $w=-1/2$, and at the two roots
 $w_1$, $w_2$ of $1+3w+4w^2=0$, we remark that the solution carrying a logarithmic term is
in fact a solution of $\, Z_2 \cdot N_1$.
Therefore, the three solutions of the differential operator 
$Y_3 \cdot Z_2 \cdot N_1$, emerging from
 $Y_3$, are analytical at the non physical singular points $w=1$, $w=-1/2$,
and at the quadratic roots of $1+3w+4w^2=0$. At the
singular point $w=1/4$, we also note that the differential operator $Z_2 \cdot N_1$ 
is responsible of the $\, (1-4w)^{-1}$ behavior. We will  then expect the
"ferromagnetic constant" $I_3^{+}$ to be localized in the blocks of the
connection matrix corresponding to the solutions of 
the order three differential operator $Z_2 \cdot N_1$ at the point $w=1/4$.

\vskip 0.2cm

\centerline{
\begin{tabular}{|l|l|l|l|l|l|l|}
\hline
&  &  &  &&&  \\ 
$w$-singularity & $Z_2\cdot N_1$ & $N$ & $P$  & $Y_3\cdot Z_2\cdot N_1$ & $N$ & $P$
\\ 
&  &  &  &&&  \\ 
\hline
&  &  &  &&&  \\
$0$ & $2,1,1$  & $1$ & $1$ & $3,2,2,1,1,1$ & $3$ & $2$ \\ 
&  &  &  &&&  \\ 
$-1/4$ & $1,0,-1/2$  & $0$ & $0$&  $2,1,0,0,0,-1/2$ & $2$ & $2$\\ 
&  &  &  &&&  \\ 
$1/4$ & $-1,-1,-3/2$  & $1$ &$1$  & $0,0,0,-1,-1,-3/2$ & $3$ & $2$\\
&  &  &  &&&  \\ 
$\infty $ & $1,0,0$ & $1$ &$1$& $2,1,1,1,0,0$& $3$ &$2$ \\ 
&  &  &  &&&  \\ 
$-1/2$ & $3,1,0$   & $1$ &$1$  & $4,3,3,2,1,0$& $1$ & $1$\\ 
&  &  &  &&&  \\ 
$1$ & $3,1,0$   & $1$ &$1$  & $4,3,3,2,1,0$ & $1$ &$1$\\ 
&  &  &  &&&  \\ 
$\frac{-3\pm i\sqrt{7}}{8}$ & $1,1,0$  & $1$ &$1$& $4,3,2,1,1,0$ & $1$ & $1$ \\ 
&  &  &  &&&  \\ \hline
\end{tabular}
}

\vskip 0.2cm
\textbf{Table 1:} Critical exponents for each regular singular point for
the differential operators $Z_2\cdot N_1$ and $Y_3 \cdot Z_2 \cdot N_1$.
The columns $N$ show the number of solutions with logarithmic terms.
The columns $P$ show the maximum power of the logarithm occurring in the
solutions.

\vskip 0.2cm

As far as explicit calculations are concerned,
a well-suited  basis necessary for explicitely writing
 connection matrices exists and can be described.
Considering the order six operator $\, L_6=\, Y_3 \cdot Z_2 \cdot N_1$, we construct
the local solutions, sequentially, as the global solution of $N_1$ then
the two solutions coming from $Z_2\cdot N_1$, 
to which we add the three further solutions coming from
$Y_3 \cdot Z_2 \cdot N_1$.
We will use below this well-suited bases.

\section{Connection matrices for $\tilde{\chi}^{(3)}$ }
\label{con-mat}
Using a very simple method, let us show, in the case
where one has an exact Fuchsian differential equation, 
 that {\em one can actually} 
very simply, {\em and very efficiently}, obtain
the connection matrices between two sets of series-solutions valid at
two different points. The method consists in equating, at some matching
points, the two sets of series corresponding, respectively, to expansions
around $w=0$ and, for instance, $w=1/4$.
The matching point should be in the radius of convergence of both series.
The singular points (i.e., $w=0$ and $w=1/4$) should be neighbors, having
no other singularity in between.
Recall that the differential equation for $\tilde{\chi}^{(3)}$ has eight
regular singular points, the point at infinity, five on the real axis and
two ($w_1$ and $w_2$) on the upper and lower half plane each.
At a given singular point $w_s$, the
solutions are obtained as series in the variable $x$, where $x=w$
(resp. $x=1/w$) for the point $w_s=0$ (resp. $w_s=\infty$)
and $x=1-w/w_s$ for the other regular singular points. We take the definition
$\ln(x)=\ln(-x)+i\,\pi$ for negative values of $x$ which corresponds to
matching points in the lower (resp. upper) half-plane for $w>0$
(resp. $w<0$).

The computation of the connection matrix should be more efficient when
two ``neighboring'' singularities are, as far as possible, far away
from the other singularities and, especially, when the test points
chosen half-way are, as far as possible,  far from the other
singularities, in order not to be ``polluted'' by the other singularities. 
We remark that one can calculate, in this way, just ``neighboring''
singularities: connection matrices of 
two singularities $w_1$, $\, w_r$ that are
 not ``neighbors'' should be deduced 
using some path of ``neighboring'' connection matrices:
\begin{eqnarray}
\label{neighbors}
C(w_1, \, w_r) \, = \, C(w_1, \, w_2) 
\cdot C(w_2, \, w_3) \, \cdots\, C(w_{r-1}, \, w_r) 
\end{eqnarray}
This is the prescription we take for the singular points on the
real axis and the singularity $w_1$ lying in the upper half-plane.
For the singularity $w_2$ lying in the lower half-plane,
the connection matrix is calculated from:
\begin{eqnarray}
\label{neighbors-w2}
C(0, \, \, w_2) \, = \, C^{*}(0, \, -1/4) \cdot C^{*}(-1/4, \, w_1) \,=\,
C^{*}(0, \, \, w_1)
\end{eqnarray}
where $^{*}$ denotes the complex conjugate.

Let us remark that changing the variable $\, w$ we are working with,
to the more traditional $\, s \, = \, \sinh(2K)$ variable, 
or the usual high-temperature (resp. low temperature) variable 
$\, t \, = \, \tanh(K)$, or the variable
 $\, \tau\, = \, (1/s-s)/2$, modifies the distribution
of singularities in the complex plane and their radii of convergence.
However, the method can still be used.
One can use that freedom in the choice of the
expansion variable to actually improve the convergence of our 
calculations.

\subsection{Connecting solutions}
\label{z2n1}
Let us first show, as an example, how we compute 
the connection matrix between
two neighboring regular singular points ($w=0$ and $w=1/4$)
for order three differential operator $\, Z_2 \cdot N_1$.
Around the singular point $w=0$, the local solutions are
two Frobenius series (one being the global solution ${\cal S}(N_1)$) and
a series with a logarithmic term. The chosen basis is then (where $x=w$):
\begin{eqnarray}
\label{sol12w0}
S_1^{(0)} (x) &=&\, \, {\cal S}(N_1)(x), \qquad   
S_2^{(0)}(x)  =  [0, 1, 5, 26, 106, 484, \cdots ],       \\
\label{sol3w0}
S_3^{(0)}(x)  &=&\,\,   S_{2}^{(0)}(x) \cdot \ln(x) + S_{30}^{(0)}(x) 
\end{eqnarray}
with:
\begin{eqnarray}
\label{sol3nonF}
S_{30}^{(0)} (x) &=& [0, 0, 0, 6, 26, 529/3, 2149/3, \cdots ]   
\end{eqnarray}
where $\, [a_0, \, a_1, \, a_2, \, \cdots, ]$ denotes the series
$\, a_0+ \, a_1\, x+ \, a_2\, x^2+  \, \cdots$  
There are three independent series $S_1^{(0)}$, $S_2^{(0)}$ and $S_{30}^{(0)}$,
since the operator $Z_2 \cdot N_1$ is of order three.
Similarly, around $w=1/4$, the local solutions read (with $x=1-4w$ and,
where again, $S_1^{(1/4)}$ is the global solution corresponding to operator $N_1$):
\begin{eqnarray}
\label{solz2n114}
S_1^{(1/4)}(x)  &=& \, \,  {\cal S}(N_1)(x),    \\
\label{solz2n114-2}
S_2^{(1/4)}(x)  &=& \, \, 
{{1} \over {x}} \, -{{3} \over {4}}\,  -{{5} \over {96}} \cdot x \, 
 -{{3} \over {64}} \cdot x^2  \, -{{1801} \over {55296}} \cdot x^3 
 \, +\, \cdots      \\
\label{solz2n114-3}
S_3^{(1/4)}(x)  &=&\,\,   S_{2}^{(1/4)}(x) \cdot \ln(x) \, +S_{30}^{(1/4)}(x)
\end{eqnarray}
with:
\begin{eqnarray}
\label{sol3w14nonF}
S_{30}^{(1/4)} (x) &=&\,  [3/8, -367/5760, -193/6720, -244483/6635520, \cdots ]
\end{eqnarray}

The series $S_i^{(0)}$ are defined around $w=0$, and are convergent in a
radius of $1/4$, which corresponds to the nearest regular singular
point (i.e., $w=1/4$). Similarly, the solutions $S_i^{(1/4)}$ are convergent
in the disk centered at $w=1/4$ with same radius (i.e., $1/4$).
Between the points $w=0$ and $w=1/4$, there is a region where both
sets of solutions ($S_i^{(0)}$ and $S_i^{(1/4)}$) are convergent. This region
corresponds to the common area between two disks centered respectively
at $w=0$, and $w=1/4$, with the same radius $1/4$.

Connecting the local series-solutions at the regular singular points $w=0$,
and $w=1/4$, amounts to finding the $3 \times 3$ matrix  $C(0,1/4)$
such that
\begin{eqnarray}
\label{Cz2n1}
S^{(0)} \,  =\,\,   C(0,1/4) \cdot S^{(1/4)}
\end{eqnarray}
where $\, S^{(0)} $ (resp. $\,S^{(1/4)}$) denotes 
the vector with entries $S_i^{(0)}$ (resp.  $S_i^{(1/4)}$).
The solutions $S_i^{(0)}$ and $S_i^{(1/4)}$ are evaluated at three
arbitrary points around a point $\, x_c$
belonging to both convergence disks of the series-solutions $S_i^{(0)}$
and $S_i^{(1/4)}$.

Equation (\ref{Cz2n1}) is thus a linear system of nine unknowns. The
entries of the connection matrix $C(0,1/4)$ are obtained in floating
point form with
a large number of digits. These entries are ``recognized'' in symbolic
form and matrix $\, C(0,1/4)$ then reads:
\begin{eqnarray}
C(0,1/4) \, = \, \, \, 
\left [\begin {array}{ccc}
1&0&0\\
\noalign{\medskip}
1&-{\frac {9\sqrt{3}}{64\pi}}\,\left( {{2}\over{3}}-\ln(24) \right)  &
-{\frac {9\sqrt{3}}{64\pi}} \\
\noalign{\medskip}
0&-{\frac {3 \pi \sqrt{3}}{32}} & 0
\end {array}\right]
\end{eqnarray}

The entries of this matrix are combinations of radicals, of powers
of $\pi$ and logarithms of integers. 
Note that there is no straightforward manner to recognize
numerical values such as the ones displayed above.
However, it is possible, in a ``tricky way'', to get
rid of the logarithms of integers in the entries, and  obtain as many
zero entries as possible.
This is shown, in the following, for this very example.

The series, in the set of local solutions $S_i^{(1/4)}$, are solutions of
the differential equation (ODE) corresponding to the third order differential
operator $\, Z_2\cdot N_1$ at the regular singular point $w=1/4$.
It is obvious that any linear combination of these series is also a solution
of the differential equation.
Consider the following combination instead of the third component in
(\ref{solz2n114-3}):
\begin{eqnarray}
\label{sol3w14}
S_3^{(1/4)}(x)\, \,\, \,  \longrightarrow \,\,  \,\,\, \, 
 \left( \ln(x/24)+2/3 \right)\cdot  S_2^{(1/4)}(x) \,  +S_{30}^{(1/4)}(x)
\end{eqnarray}
By writing the argument of the logarithm as $x/24$, there will be no
logarithm in the connection matrix. Furthermore, by adding the second
component of the basis to the third component with a factor of $2/3$,
the entry $(2,2)$ of the connection matrix will be canceled.
The connection matrix then reads:
\begin{eqnarray}
C(0,1/4) \, = \, \,\, \,
\left [\begin {array}{ccc}
1&0&0\\
\noalign{\medskip}
1&0&-{\frac {9}{64}}\,{\frac {\sqrt {3}}{\pi }}\\
\noalign{\medskip}
0&-{\frac {3 \pi \sqrt {3} }{32 }}&0
\end {array}\right]
\end{eqnarray}

These tricks, based on well chosen linear combinations of the solutions, 
allow us to obtain as many zeroes as possible,
and to get rid of the logarithms. They will be used
in order to compute the connection matrix for $L_6$
between the point $w=0$ and, respectively, $\, w=1/4$,
$\, w=-1/4$ and $\, w=\infty$.

The chosen well-suited basis of solutions, at each regular singular point
calls for some comment. The factorization of the differential operator $L_6$
being $\, Y_3 \cdot Z_2 \cdot N_1$, our method of producing the solutions,
sequentially, allows one to determine from which differential operator a given
solution emerges.
Near the points $w=0$, $w=\pm 1/4$, and $w=\infty$,
the third order differential operator $Y_3$ brings three
solutions (see Table 1), one
Frobenius series, one solution with a $\, \log$ term, and one solution
with a $\, \log^2$ term, denoted respectively $\tilde{S}_4$, $\tilde{S}_5$
and $\tilde{S}_6$.
The solutions of the differential operator $Y_3$ itself are of elliptic
integral type (see Appendix B).
These elliptic integrals behave around $w=\pm 1/4$ (resp. $w=\infty$)
like $g(t)\cdot \ln(t/16) +f(t)$, with $t=1-16w^2$ (resp. $t=1/16w^2$),
$g(t)$ and $f(t)$ being series with rational coefficients.
One may then assume that the logarithmic term that appears
in the solutions of $L_6$, inherited from $Y_3$, 
will be of the form $\, \ln((1-16w^2)/16)$, near $w=\pm 1/4$, and of the
form $\, \ln(1/256/w^2)$, near $w=\infty$.
The general form of combination for the fourth to sixth components of
the well-suited basis will be:
\begin{eqnarray}
\label{combination}
\tilde{S}_4 & \longrightarrow & \tilde{S}_4 \nonumber \\
\tilde{S}_5 & \longrightarrow & \tilde{S}_5 +\left(a_1-\ln(c) \right) \cdot \tilde{S}_4  \\
\tilde{S}_6 & \longrightarrow & \tilde{S}_6 +2\left(a_1-\ln(c) \right) \cdot \tilde{S}_5
+\left(\ln(c)^2-2a_1\ln(c)+a_2 \right) \cdot \tilde{S}_4 \nonumber 
\end{eqnarray}
where $c=1, \, 8, \, 16$ for the basis at, respectively, $w=0$, $w=\pm 1/4$
and $w=\infty$.
The values of the parameters $a_1$ and $a_2$ depend on each basis.

Note that the argument in $\ln(x/24)$ in the series solutions of the differential
operator $Z_2 \cdot N_1$ at $w=1/4$ will be $\ln(x/4)$ and $\ln(x/24)$
at respectively $w=\infty$ and $w=1$. Similarly to $Y_3$, these arguments
may come from the explicit solutions of $Z_2$.

\subsection{Connection matrix between $w=0$ and $w=1/4$}
The first three local solutions at $w=0$
are given by (\ref{sol12w0}), (\ref{sol3w0}), (\ref{sol3nonF}),
and the fourth, fifth and sixth solutions read
\begin{eqnarray}
S_4^{(0)}(x) &=&  [0, 1, 9, 34, 178, 692, \cdots \, ],            \nonumber \\
S_5^{(0)}(x) &=& \,\, S_{4}^{(0)}(x) \cdot \ln(x)
 + S_{50}^{(0)}(x)   - S_{4}^{(0)}(x)/4, \nonumber \\
S_6^{(0)}(x) &=& \,\,  S_{4}^{(0)}(x) \cdot \ln^2(x)\, 
+  2 \, \bigl( S_{50}^{(0)}(x) - S_{4}^{(0)}(x)/4 \bigr) \cdot \ln(x)  \nonumber \\
&&  \,\, + S_{60}^{(0)}(x) - S_{50}^{(0)}(x)/2  +25 \, S_{4}^{(0)}(x)/16
\nonumber
\end{eqnarray}
with:
\begin{eqnarray}
S_{50}^{(0)}(x) &=& [0, 0, 0, -2, 34, 241/3, \cdots \,  ],  \nonumber \\
S_{60}^{(0)}(x) &=& [0, 0, 0, 0, -19/3, -7693/72, -575593/1800, \cdots \, ].  \nonumber
\end{eqnarray}
At the singular point $w=1/4$, we make use of the combination
(\ref{combination}) which amounts to taking $x/8$ as argument of the
logarithms in the fourth, fifth and sixth component.
The parameters $a_1$ and $a_2$ in (\ref{combination}) are respectively $23/6$ and $41/9$.
The first three local series at $x=1-4w$ are given in
(\ref{solz2n114}), (\ref{solz2n114-2}), (\ref{sol3w14nonF}), (\ref{sol3w14}), and
the fourth, fifth and sixth read
\begin{eqnarray}
\label{basis14}
S_4^{(1/4)} (x) &=&\,\,   [1, -1/8, 3/16, 29/512, \cdots ],         \\
S_5^{(1/4)} (x) &=&\,\,  (\ln(x/8) + 23/6) \cdot S_{4}^{(1/4)}(x)+ S_{50}^{(1/4)}(x), \nonumber \\
S_6^{(1/4)} (x) &=&\,\,   \left( \ln^2(x/8)  
+{{23}\over{3}} \ln(x/8) +41/9 \right) \cdot  S_{4}^{(1/4)}(x)    \nonumber  \\
&&  \qquad     \quad        +  2\, \left( \ln(x/8) + 23/6 \right)\cdot  S_{50}^{(1/4)}(x)  
 + S_{60}^{(1/4)}(x)\nonumber
\end{eqnarray}
with:
\begin{eqnarray}
S_{50}^{(1/4)} (x) &=& \,\,  [0, 457/480, -2231/1680, -128969/184320, \cdots \,  ]    \nonumber \\
S_{60}^{(1/4)} (x) &=& \,\,  [0, -967/100, 4312219/470400, 595578701/116121600,  \cdots \, ] \nonumber
\end{eqnarray}

Connecting both solutions amounts to solving a linear system of 36 unknowns
(the entries of the connection matrix).
We have been able to recognize these entries which are obtained
in floating point form with a large number of digits.
The connection matrix $\, C(0,1/4)$
for the order six differential operator $L_6$ reads:
\begin{eqnarray}
\label{C014}
&& C(0,\, 1/4)\, = \, \, \\
&& \left[ \begin {array}{cccccc} 
1&0&0&0&0&0\\
\noalign{\medskip}
1&0&-{\frac {9 \sqrt {3}}{64 \pi}}&0&0&0\\
\noalign{\medskip}
0 &-{\frac {3 \pi \sqrt {3}}{32}} &0&0&0&0\\
\noalign{\medskip}
5&{\frac {1}{3}}-2 \cdot I_3^{+}&{\frac {3 \sqrt {3}}{64 \pi}}&0&0&{ \frac {1}{16 \pi^2} }\\
\noalign{\medskip}
-{\frac{5}{4}}&-{\frac {3 \pi \sqrt {3}}{32}} &{\frac {45 \sqrt {3}}{256 \pi}}&0&{\frac{1}{32}}&0\\
\noalign{\medskip}
{\frac{29}{16}}-{\frac{2 \pi^2}{3}}&{\frac {15 \pi \sqrt {3}}{64}}
&-{\frac {225 \sqrt {3}}{1024 \pi}} -{\frac {3 \pi \sqrt {3}}{64}}&{\frac {\pi^2}{64}}&0&0
\end {array} \right]
\nonumber
\end{eqnarray}

\vskip 0.2cm
Some comments on how these entries have been ``recognized'' will be given
below.
Let us remark that, once the entries of the connection matrix have
been obtained, 
a further change of basis can be made
to get it as "simple" as possible.

\subsection{Connection matrices between $w=0$ and the other
regular singular points}
\label{connzeroregul}
The chosen basis of solutions and the connection matrices between
$w=0$  (high or low temperature) and, respectively, the
 anti-ferromagnetic point $\, w=-1/4$
and the point $w=\infty$ (corresponding to $s=\pm i$) are
given in Appendix C.

The chosen basis, used for the regular singular points 
$w=1, -1/2$ and $1+3w+4w^2=0$, are given in Appendix D together with the
corresponding connection matrices with the point $w=0$.
Many entries are ``recognized'' and, in particular, {\em those required to find
the singular behavior of the physical solution}.
They correspond to the third column of
matrices given in Appendix D.

The connection matrix between each pair of neighboring singular points is
computed with the well defined procedure described above.
The connection matrix between $w=0$, and a non neighbor
singular point, is computed using (\ref{neighbors}). For instance,
$C(0,1)$ is computed from $C(0,1/4)$ and $C(1/4,1)$ as
$C(0,1) \, = \, C(0,1/4) \cdot C(1/4, 1)$
which says that the solutions defined at $w=1/4$ connected to the solutions
defined at $w=0$, are also the solutions that are connected to the
solutions defined at $w=1$. 

To be more confident of this prescription, let us underline that
the connection
matrices $C(0,1)$ and $C(0, -1/2)$, deduced from (\ref{neighbors}),
will be used below to confirm known dominant singular behavior of
$\tilde{\chi}^{(3)}$ and find the subdominant behavior.

\subsection{Comments and remarks}
\label{comments}
The connection matrices between $w=0$ and the other singular points
are structured in blocks.
The latter, due to the factorization of the differential operators
and to the sequential building of
the solutions, are easily recognized.
The block $(1,2,3) \times (1,2,3)$ is associated with the third order differential
operator $Z_2 \cdot N_1$.
The block $(4,5,6) \times (4,5,6)$ represents the connection between
the solutions (at both $w=0$ and the other singular points being considered)
of $L_6$ that {\em are not} solutions of $Z_2\cdot N_1$.
The "ferromagnetic constant" $I_3^{+}$ appears in the connection matrix
between $w=0$ and $w=1/4$, as mentioned earlier,
in the block $(1,2,3) \times (1,2,3)$ at the column
corresponding to the  $S_2^{(1/4)}$ (see (\ref{solz2n114-2})) solution of
the third order differential operator $Z_2 \cdot N_1$.

To compute the connection matrix, we have used the differential operator
$L_6$ which has a unique factorization. If, instead, we consider the
differential operator $L_7$, the next solution (around $w=0$),
that comes from $\, M_1$, will
be the series (\ref{singledout}) and will correspond to $\tilde{\chi}^{(3)}$.
This seventh solution is expressed as a linear combination of the already existing
components and of the solution of the differential operator $L_1$. We can then choose
 to add the latter as the seventh solution. The connection matrix will
have a $1$ at the entry $(7,7)$ and zero elsewhere on the seventh line (and
column), since the solution of the differential operator $L_1$ is global.
By considering another factorization of $L_7$, we will get
the same structure with an obvious relabelling.

Let us make a few computational remarks on the calculation of
these connection matrices.
At the matching of the series-solutions for which $1500$
coefficients\footnote[3]{For some checks, $3000$ terms have been generated.}
are generated from homogeneous and non-homogeneous recurrences,
the entries of the matrix are computed with
$\, 800$ digits for all the singular points. The numbers that come in
floating form are ``recognized'' as powers of $\pi$, radicals and rational
numbers, and are in agreement up to $400$ digits\footnote[4]{Let us note 
that the "ferromagnetic constant" $I_3^{+}$ has been obtained
up to more than 400 exact digits.}
for the connection between the solutions
at $w=0$ and $w=\pm 1/4$, and up to $100$ digits for the connection involving
other singular points like, $w=1$. This fact is related to the convergence
rate of the series at the (midway) chosen matching points.
For instance, between $w=0$ and $w=1/4$, the matching points near $w=1/8$
are such that both series (at $w=0$, and $w=1/4$), which have the same
radius of convergence, will be faithfully reproduced with the number
of terms used in the series.
The matching of the solutions between $w=1/4$, and $w=1$, will then require
more terms to fulfill the same accuracy than in the $(w=0)$-$(w=1/4)$ situation.
This is due to the fact that, at $w=1$, the convergence radius being $3/4$,
the matching points, which should be in the common region of both disks,
are closer to $w=1/4$ than to $w=1$.
As a general rule, the matching points are chosen around the middle of
the segment in the common area between the convergence  disks of
the two regular singular points for which the connection matrix is computed.

The difficulty in finding ``non-local'' connection matrices is rooted in the recognition
of the entries. We have given the connection matrix between $w=0$ and $w=1/4$ with
entries fully recognized (apart from $I_3^{+}$) to show that the method actually works
and is efficient. For the matrices concerning the connection between $w=0$ and the 
other singular points, we have concentrated our effort {\em on the entries that will
show up in the physical solution}.
We should note that there is no reason to expect the other (not yet recognized) 
entries to be "simply" combinations of $\pi$'s, $\log$'s and radicals.
These entries are probably  {\em valuations of holonomic functions}.
This was clearly seen in numerous examples we tackled of various differential
equations (of order two and three) with known solutions
of hypergeometric type.
The recognition process used the fact that {\em we actually found the explicit solutions of
differential operator} $Y_3$ and, thus, knew how the numerical logarithms
can be tackled.
These were  ``absorbed'' in the basis. We know, on the other hand, that the problem
is roomed with hypergeometric functions. We then expect some $\pi$'s to be
present. For the entries consisting of simple product expression, recognizing
the number amounts to performing simple arithmetic operations.
Note that considering the inverse of the connection matrix,
some entries also show up as simple rationals. The combination where $\pi$'s,
radicals
and rationals appear additively comes from looking to, for instance, at the
determinant of the matrices, or block matrices, which happen to be easily
recognizable (in fact rational or quadratic numbers for the roots of $1+3w+4w^2=0$).

Another remark is the following. We first obtained the connection matrix
(\ref{C014}) in some general basis. The matrix had more non zero
entries compared to (\ref{C014}) involving powers of $\pi$, radicals and
also $\ln(3)$ and powers of $\ln(2)$. The well-suited basis we chose
has "evacuated" all these log's in the entries of the matrix, lessening the 
recognition-process effort. But, of course, all these logs will reappear in the final
result such as the singular behavior of the physical solution as next sections
will show.

\section{The physical solution $\, \tilde{\chi}^{(3)}$ and its singular behavior}
\label{from}
The calculations of connection matrices are obtained
straightforwardly from the well-defined numerical process described
in Section \ref{con-mat}. Having $\, N$ singularities,
one needs $\, N-1$ such connection matrices in 
order to find the correspondence between all these well-suited bases
of series-solutions.

Let us focus on some particular entries of these
 various connection matrices, namely
the entries corresponding to the decomposition of $\, \tilde{\chi}^{(3)}$
in terms of the various well-suited bases associated with
 each singularity.
We have used the fact that the physical solution (corresponding
to $\, \tilde{\chi}^{(3)}$) decomposes as the solution of differential operator $L_1$,
${\cal S}(L_1)$ (which is $\tilde{\chi}^{(1)}/2$) and the physical solution of
the operator $L_6$ denoted $\Phi_6(w)$ \cite{ze-bo-ha-ma-04,ze-bo-ha-ma-05}:
\begin{eqnarray}
\tilde{\chi}^{(3)} (w)  \, \, = \, \,  
{\frac{1}{6}} \, \tilde{\chi}^{(1)} \, + \, \Phi_6(w) \nonumber 
\end{eqnarray}
Furthermore, our well-suited basis of solutions at the singular point $w=0$,
does not contain, as a component,
the physical solution $\Phi_6(w)$ which is given in terms of the previously considered
components as:
\begin{eqnarray}
\Phi_6(w) \, = \,\, {\frac{4}{3}}\, S_1^{(0)} \,
- {\frac{1}{12}}\, S_2^{(0)}\,
- {\frac{1}{4}}\, S_4^{(0)}
\end{eqnarray}

This physical solution can now be easily obtained
from the connection matrices between $w=0$ and any regular singular
point, that we denote $w=w_s$ (with $x=w$, $x=1/w$ for
respectively $w=0$ and $w=\infty$ and $x=1-w/w_s$, otherwise) as:
\begin{eqnarray}
\Phi_6(x) \, = \, \sum_{j=1}^6 \,
\Bigl( {\frac{4}{3}}\, C(0, w_s)_{1j} 
- {\frac{1}{12}}\, C(0, w_s)_{2j} -
{\frac{1}{4}}\, C(0, w_s)_{4j} \Bigr) \cdot  S_j^{(w_s)} \nonumber 
\end{eqnarray}
For instance, at the ferromagnetic critical point,
this physical solution $\, \Phi_6(x)$ can easily be deduced from (\ref{C014}),
and written as:
\begin{eqnarray}
\Phi_6(x) \, = \,\,\, -{\frac{1}{4}}\, \Bigl({{1}\over{3}}
-2\,I_3^{+} \Bigr) \cdot  S_2^{(1/4)}
 \,- \,{\frac{1}{64\,\pi^2}}\, S_6^{(1/4)} \nonumber
\end{eqnarray}
$S_2^{(1/4)}$ and $S_6^{(1/4)}$ are known from their series expansion
(\ref{solz2n114-2}), (\ref{basis14}).
This equation, giving the full expansion of
$\tilde{\chi}^{(3)}$ at $w=1/4$, can hardly be obtained directly from
the integrals defining $\tilde{\chi}^{(3)}(w)$. One has similar expansions
for all the other singular points.

\subsection{Singular behavior of $\tilde{\chi}^{(3)}$}
\label{singbev}
Knowing the
behavior of solutions $S_j^{(w_s)}$ near each regular
singular point,
it is straightforward to get the singular behavior
at those points for the physical solution $\Phi_6$ (and thus
$\tilde{\chi}^{(3)}$).

Considering the critical behavior of  $\tilde{\chi}^{(3)}$
near the ferromagnetic critical point $\, w \, = \, 1/4$, and
denoting $\, x\, = \, 1\, -4\, w$,
the singular part of the ``physical'' solution  
$\tilde{\chi}^{(3)}$ reads:
\begin{eqnarray}
\label{singpart-f}
\tilde{\chi}^{(3)} ({\rm singular},1/4) \,\, &  = & \,\, \,\, \,
{\frac{1}{2}} \,\, \, {{I_3^{+}} \over {x}} \, \, 
-   {\frac{1}{64\,\pi^2}}\, S_4^{(1/4)}\cdot  \ln^2(x)   \\
&&+  {\frac{1}{32\,\pi^2}} \Bigl( (3\, \ln(2)
 -{\frac{23}{6}})\cdot  S_4^{(1/4)}- S_{50}^{(1/4)} \Bigr)\cdot \ln(x)  \nonumber 
\end{eqnarray}
where  $\, I_3^{+}$ {\em is actually the "ferromagnetic constant"} (\ref{I3p}),
and $S_i^{(1/4)}$ the series defined in the well-suited basis (\ref{basis14})
 at $w=1/4$.
The results agree with previous results of B. Nickel,
but the correction
terms are new\footnote[2]{These results
have also been found by B. Nickel (private communication).},
in particular
the term $\, 3 \ln(2)/32/\pi^2$ in (\ref{singpart-f}).  
In terms of the $\, \tau\, = (1/s-s)/2 $ variable introduced 
in~\cite{nickel-99,or-ni-gu-pe-01b,or-ni-gu-pe-01},
the singular part (\ref{singpart-f})
reads:
\begin{eqnarray}
\tilde{\chi}^{(3)}({\rm singular},\tau \simeq 0) \, \simeq \,\, 
{{I_3^{+}} \over {  \tau^2}}\,  \,
- {{ \ln^2(\tau)} \over {16\, \pi^2}}\, \,
 +  \left( \ln(2) \,-{{23}\over{24}} \right)\cdot  {\frac{\ln(\tau)}{4\,\pi^2}}
+ \cdots   \nonumber
\end{eqnarray}

Near the antiferromagnetic critical point $\, w \, = \, -1/4$,
$\tilde{\chi}^{(3)}$ behaves as:
\begin{eqnarray}
\label{singpart-af}
\tilde{\chi}^{(3)} ({\rm singular},-1/4) \,  & = & \,\,
 - {\frac{1}{32\pi^2}} \, S_4^{(-1/4)} \cdot  \ln^2(x) \\
&& -{\frac{1}{16\pi^2}} \, \Bigl( 3\,  (2-\ln(2)) \cdot S_4^{(-1/4)} \,+
 S_{50}^{(-1/4)} \Bigr)\cdot \ln(x)              \nonumber 
\end{eqnarray}

At the non-physical singularities $w=1$ and $w=-1/2$
the physical solution behaves, respectively, like:
\begin{eqnarray}
\tilde{\chi}^{(3)} ({\rm singular}, 1) \,  \,   =  \,\,\,  \, \, 
{\frac{\sqrt {3}}{27\, \pi}} \cdot  S_2^{(1)} \cdot \ln(x) 
\end{eqnarray}
and
\begin{eqnarray}
\tilde{\chi}^{(3)} ({\rm singular}, -1/2) \, \,    =  \,\, \, \, \, 
-{\frac{8\sqrt{3}}{27\,\pi}} \cdot  S_2^{(-1/2)}\cdot  \ln(x) 
\end{eqnarray}
confirming Nickel's calculations given in \cite{nickel-00}.

At the point $w=\infty$, corresponding to the non physical singularities
$s =\pm i$, the singular behavior reads:
\begin{eqnarray}
&& \tilde{\chi}^{(3)} ({\rm singular}, \infty) \,   =  \,
 -{\frac{1}{16\pi^2}}\, S_4^{(\infty)} \cdot  \ln^2(x)  \\
&&\quad  -  {\frac{1}{8\pi^2}}\, \Bigl((4-2\pi\,i)\cdot  S_2^{(\infty)}
-(5+4\ln(2)+i\,{{\pi}\over{2}})\cdot  S_4^{(\infty)}+S_{50}^{(\infty)}
 \Bigr)\cdot  \ln(x) \nonumber 
\end{eqnarray}

At the new singularities found in \cite{ze-bo-ha-ma-04}, namely
the roots of $\, 1+3w+4w^2=0$,
 which are regular singular points of the
differential equation, the singular part of the physical solution reads,
at first sight:
\begin{eqnarray}
\tilde{\chi}^{(3)} ({\rm singular}, w_1) \, \,    =  \,\, \, 
-{{1}\over{12}} \Bigl( a_{23} +3\,a_{43} \Bigr) \cdot  S_2^{(w_1)} \cdot  \ln(x) \nonumber
\end{eqnarray}
The entries $a_{23}$ and $a_{43}$ (see the connection matrix for these points
in Appendix D) are however such that $a_{23}+3a_{43}=0$.
The {\em physical solution is thus, not singular},
at the newly found quadratic singularities, confirming our conclusion
given in \cite{ze-bo-ha-ma-05} from series analysis.

\subsection{Asymptotic series analysis}
\label{asymptoseries}
As the physical solution $\tilde{\chi}^{(3)}$ is given as a series
around $w=0$, the coefficients of the latter are controlled by the
nearest singular points (i.e. $w=\pm 1/4$).
Since the singular parts at the ferromagnetic and anti-ferromagnetic critical
points (\ref{singpart-f}), (\ref{singpart-af}) are obtained,
it is straightforward to deduce the behavior of the coefficients of series
(\ref{singledout}) for large values of $n$.
Standard study of the asymptotic behavior of the coefficients via their linear
recursion relation can be used (see \cite{wim-zei-85}). 
For our purpose, we use 
the following identity for $\ln^2(1-x)$ (where $x$ stands for $x=4w$):
\begin{eqnarray}
\label{coflog2}
&&\ln^2(1-x) \, = \, \, \,\sum_{n=2}^{\infty} \, b(n)\cdot  x^n, 
\qquad \quad \hbox{where:} \nonumber \\
&&b(n) \, = \, \,\, \sum_{i =1}^{n-1}\,  {{1} \over {i\, (n-i)}}\, 
=\, \,\, {{2} \over {n}} \cdot \Bigl(\Psi(n) \, + \gamma \Bigr)
\end{eqnarray}
where $\gamma=0.57721566 \cdots \, $ denotes Euler's constant, 
and $\, \Psi$ denotes the logarithmic derivative of the $\, \Gamma$
function. Recalling  the asymptotic expansion of $\Psi(n)$ up to $1/n^2$ for
large values of $n$, one obtains:
\begin{eqnarray}
 b(n)  \quad \rightarrow \quad
 {{2} \over {n}} \cdot \Bigl(\gamma\,
 +\ln(n) -{{1}\over{2n}}-{{1}\over{12n^2}} +\cdots  \Bigr)
\nonumber
\end{eqnarray}
With the same manipulations of $\,\ln^2(1+x)$, and inserting in
(\ref{singpart-f}), (\ref{singpart-af}), one obtains the asymptotic form 
of coefficients of $\tilde{\chi}^{(3)}/8w^9$ as:
\begin{eqnarray}
2^{-15} \cdot {\frac{c(n)}{4^n}} \, &\simeq& \,\,\,  \,  {{ I_3^{+}} \over { 2}}\,\,
-{{1} \over { 16 \pi^2}} \Bigl( {{1}\over{2}}\,+(-1)^n \Bigr)\,
 \Bigl( {{\ln(n) } \over {n}}+{{b_1}\over{n}} - {\frac{1}{2\,n^2}} \Bigr)  \nonumber \\
&& \quad  + {{1}\over{16\pi^2}} \Bigl( {{23}\over{12}}+6\, (-1)^n \Bigr)\, {{1}\over {n}}+\, 
\cdots \nonumber 
\end{eqnarray}
where $b_1 = \gamma +3\ln(2)$.

It is this parity effect in the asymptotic behavior of the coefficients
that we saw, numerically, 
(see equations (33) in~\cite{ze-bo-ha-ma-05}) where we obtained, around
$\, n \, \simeq 500$,  $c(n) \, \simeq \, 13.5 \times 4^n$ for $\, n$
even and  $c(n) \, \simeq \, 11 \times 4^n$ for $\, n$ odd.
For very large values of $n$, the asymptotic value of the coefficient $c(n)/4^n$
is thus $2^{14}\cdot I_3^{+} \simeq 13.34415467 \cdots $.

\section{Monodromy matrices for $\tilde{\chi}^{(3)}$}
\label{diffgalL7}
\subsection{Sketching the differential Galois group of $L_7$}
As a consequence of the direct sum (\ref{factochi3}),
the differential Galois group of $L_7$ reduces (up to a product by
$\mathcal{C}$)
to the differential Galois group of $L_6$.
From the factorization of $L_6$,
one can immediately deduce that 
the differential Galois group of $\, L_6$
is the semi-direct product of the differential Galois group
of $\, Y_3$, of the  differential Galois group of $\, Z_2$
and of the differential Galois group of $\, N_1$ (namely $\, \mathcal{C}$).

In some ``well-suited global basis'' of solutions,
the form of  the $\, 6 \times 6$ matrices
representing the differential Galois group of $\,L_6$, reads:
\begin{eqnarray}
\label{generform}
  \left[ \begin {array}{cc} 
{\bf A} &  {\bf 0}\\
\noalign{\medskip}
{\bf H}  & {\bf B}
\end {array} \right], 
\qquad {\rm with}\,\, \qquad
 {\bf A} \, = \,
 \left[ \begin {array}{cc}
b &  0 \\
\noalign{\medskip}
{\bf h} & {\bf g}
\end {array} \right]\nonumber 
\end{eqnarray}
where the  $\, 2 \times 2$ matrix $ {\bf g}$, and $\, 3 \times 3$ matrix
${\bf B}$ correspond, respectively, to the differential Galois group
of $\, Z_2$ and $\, Y_3$.
The $\, 3 \times 3$ matrix ${\bf A}$ 
is associated with the differential Galois group of $ \, Z_2 \cdot N_1$,
and the $\, 3 \times 3$ matrix ${\bf H}$ 
corresponds  to 
the fact that we have a {\em semi-direct} product 
of the differential Galois group of $\, Y_3$ and $\,  Z_2 \cdot N_1$ 
in $\, L_6 \, = \, L_3 \cdot Z_2 \, N_1$.

Many papers (for instance~\cite{BermSing,Berm,Weil1,Weil2,Weil3})
describe how to calculate the differential Galois groups of order 2 and order
3 differential operators.
The  differential Galois group
of $\, L_7$ will be deduced in a forthcoming publication~\cite{Weil}. 

To go beyond this sketchy description of the differential Galois group, one
needs to calculate specific elements 
like the monodromy matrices expressed in a common basis.

\subsection{Monodromy matrices rewritten in the $\, w=0$ basis}
\label{rewritten}
Having the connection matrices between $w=0$ and each singularity, the local
monodromy matrices expressed in their own well-suited basis of (series)
solutions, can be rewritten in a {\em unique} global basis  
valid for all singularities. This will allow us, in a second step,
to calculate their products and thus generate the differential Galois group.
Let us define the $2 \times 2$ and $3 \times 3$ matrices
\begin{eqnarray}
A =
 \left [
\begin {array}{cc}
1&0  \\
\noalign{\medskip}
\Omega&1 \\
\end{array}
\right], \quad \quad\quad 
B =
 \left [
\begin {array}{ccc}
1&0&0  \\
\noalign{\medskip}
\Omega &1 &0\\
\noalign{\medskip}
\Omega^2 & 2\Omega & 1\\
\end{array}
\right]
\end{eqnarray}
where $\, \Omega$ denotes $\, 2\, i\, \pi$ and corresponds to the
translation of the logarithm when performing a complete rotation
around the regular singular 
point: $ \ln(w) \, \rightarrow \, \ln(w) + \Omega$.

The expression of the
local monodromy matrix around each regular singular point $w_s$
in its own well-suited basis of (series) solutions reads:
\begin{eqnarray}
\label{localmono}
  {\it l}(w_s)  \, = \,      
 \left [
\begin {array}{ccc}
\epsilon & 0& 0  \\
\noalign{\medskip}
0 & C & 0 \\
\noalign{\medskip}
0 & 0 & D  \\
\end{array}
\right]  
\end{eqnarray}
where, $\epsilon$ and the $\, 2 \times 2$ blocks $C$, and  $\, 3 \times 3$ blocks
$D$, are such that:
\begin{eqnarray}
w=0, \quad w=\infty &\,\rightarrow\,& \quad\quad \epsilon=+1, \,\, C=A, \,\, D=B \nonumber \\
w=1/4,              &\,\rightarrow\,&\quad \quad\epsilon=-1, \,\, C=A, \,\, D=B \nonumber \\
w=-1/4,             &\,\rightarrow\,&\quad\quad \epsilon=-1, \,\, C=Id, \, \,\, D=B \nonumber \\
w=1,\, -1/2,\, -3/8 \pm i\,\sqrt{7}/8, &\,\rightarrow\,&\quad
\quad \epsilon=+1, \,\, C=A, \,\, D=Id \nonumber 
\end{eqnarray}

The monodromy matrix around any singularity $w=w_s$ expressed in terms of
the  $(w=0)$  well-suited basis, and denoted $\, M_{w=0}(w_s)$, reads:
\begin{eqnarray}
\label{de0a1}
M_{w=0}(w_s)\, = \, \,  \, \,
  C(0, w_s) \cdot {\it l}(w_s)(\Omega) \cdot C^{-1}(0,w_s). 
\end{eqnarray}
In order to keep track of the $\, \pi$ corresponding to the translation of
the logarithm in the local monodromy matrix $\,{\it l}(w_s)(\Omega)$,
 and the $\, \pi$'s occurring in the expression 
of the entries of the (quite involved) connection matrix  $\, C(0, w_s)$,
we will denote the latter by  $\, \alpha \, = \, 2\, i\, \pi$. 
 
Let us focus on the singular point $w=1$.
Its monodromy matrix, expressed in terms of the
$w=0$  well-suited basis, is given by (\ref{de0a1}) with
$w_s=1$, and
where the connection matrix $\,  C(0, 1)$, matching the $\, (w=1)$
well-suited basis together with the $\, (w\, =0)$ well-suited basis, is a ``quite 
involved'' matrix given in Appendix D, with entries depending on $\, \pi$'s and 
on a set of 15 constants, not yet recognized in closed
form. The monodromy $M_{w=0}(1)$ can finally be written 
as a function of only $\, \alpha$ and $\, \Omega$:
\begin{eqnarray}
\label{MalphaOmega1}
&& 8\, \alpha^2 \cdot M_{w=0}(1)(\alpha, \Omega) \, \,  \, = \, \, \\ 
&&  \left[ \begin {array}{cccccc} 
8\,{\alpha}^{2}&0&0&0&0&0
\\\noalign{\medskip}
-48\,\alpha\,\Omega &8\,{\alpha}^{2}&-48\,\Omega&0&0&0
\\\noalign{\medskip}
0&0&8\,{\alpha}^{2}&0&0&0
\\\noalign{\medskip}-
1008\,\alpha\,\Omega &0&-1008\,\Omega&8\,{\alpha}^{2}&0&0
\\\noalign{\medskip}
12\,\alpha\, \left( 5+16\,\alpha \right)\,\Omega &0
&12\, \left( 5+16\,\alpha \right)\,\Omega &0&8\,{\alpha}^{2}&0
\\\noalign{\medskip}
-\alpha\, \left( 75+44\,{\alpha}^{2} \right)\,\Omega &0
&-\left( 75+44\,{\alpha}^{2} \right)\,\Omega  &0&0&8\,{\alpha}^{2}
\end {array} \right] 
\nonumber
\end{eqnarray}

Let us give one more example corresponding to the new quadratic singularities 
$\, 1\, + 3\, w\, + 4\, w^ 2\, = \, 0$.
The monodromy matrix around 
one of the quadratic singularities $\, w\, = \, w_1$,  expressed 
in terms of the $\, (w=0)$ well-suited basis, after the conjugation 
(\ref{de0a1}), reads:
\begin{eqnarray}
\label{globsq}
8\, \alpha^2 \cdot M_{w=0}(w_1)(\alpha,\, \Omega)\,\, = 
\left[ \begin {array}{cc} 
A&0\\\noalign{\medskip}
B&C
\end {array} \right]
\end{eqnarray}
with:
\begin{eqnarray}
\left[ \begin {array}{c} 
A\\\noalign{\medskip}
B
\end {array} \right] =
 \left[ \begin {array}{ccc} 
8\,{\alpha}^{2}&0&0
\\\noalign{\medskip}
48\,\alpha\,\Omega&8\,\alpha\, \left( \alpha+6\,\Omega \right) &-144\,\Omega
\\\noalign{\medskip}
16\,\Omega\,{\alpha}^{2}&16\,\Omega\,{\alpha}^{2}&-8\,\alpha\, \left( -\alpha+6\,\Omega \right) 
\\\noalign{\medskip}
-16\,\alpha\,\Omega&-16\,\alpha\,\Omega&48\,\Omega
\\\noalign{\medskip}
4\,\alpha\, \left( 4\,\alpha-15 \right)\,\Omega
&4\,\alpha\, \left( 4\,\alpha-15 \right)\,\Omega
&-12\, \left( 4\,\alpha-15 \right)\,\Omega
\\\noalign{\medskip}
\alpha\, a &\alpha\, a  &-3\, a
\end {array} \right]
\nonumber
\end{eqnarray}
with $\, \, a=\, \left( -40\,\alpha+12\,{\alpha}^{2}+75 \right)\, \Omega$
and
$
\,\, \left[ \begin {array}{c} 
C
\end {array} \right]\,=\,\, 8\,{\alpha}^{2}\cdot {\bf Id(3\times3)}
$.

One can actually verify that the monodromy matrix around 
the other quadratic singularity $\, w\, = \, w_2$ ($w_2$ is complex 
conjugate of $\, w_1$),  expressed 
in terms of the $\, (w=0)$-well suited basis, actually identifies with (\ref{globsq})
where  $\, \alpha$ has been changed into $\, -\alpha$.

We have totally similar results for all the other (regular) singularities.
The expression of the other  monodromy matrices $\, M_{w=0}(w_s)$,
around the other (regular) singular points $\, w=w_s$,
are displayed in Appendix E.

We saw that the connection matrices depend
on $I_3^{+}$ and on ``still not yet recognized''
(probably  transcendantal) numbers, like $x_{42}$ and $y_{41}$ (for the
connection matrix between $w=0$ and $w=\infty$).
Rewriting a monodromy matrix in a unique (global) basis 
like the $\, w=0$ basis, amounts to performing conjugation,
like (\ref{de0a1}), of simple (local) monodromy matrices depending only on 
$\, \Omega$, by these quite involved connection matrices.
As a consequence, one does expect, at first sight, these monodromy matrices,
rewritten in the unique $\, w=0$ basis, to be 
dependent on the still unknown numbers.
For instance one certainly expects the monodromy matrix around $\, w\, = 1/4$
(see Appendix E)
to be expressed in terms of the transcendental number $\, I_3^{+}$,
or the monodromy matrix (\ref{MalphaOmega1}) to depend on 15 parameters.
It is worth noting that all these  matrices $\, M(w_s)$,
 expressed in the same $\, (w=0)$ well-suited basis, turn out
 to be quite simple matrices where the
{\em entries are actually rational expressions, with integer coefficients}
of  $\, \alpha$ and $\, \Omega$.
Section (\ref{comme}) gives some hints on why this is so.

The introduction of the two parameters $\, \alpha$ and $\, \Omega$
is a nice ``trick'' to track the $\pi$'s coming from the connection
matrices versus the $\pi$'s coming from the local monodromy matrices. However, one
 should keep in mind that 
$\, \alpha$ {\em is not independent of} $\, \Omega$: the ``true''
monodromy matrices are such that $\, \alpha\, = \, \Omega$ ($\Omega$
being equal to $\, 2\, i\, \pi$).
Let us denote these  ``true'' monodromy matrices
by  $\, M_i$,  $\,\, i \, = \, 1, \cdots , \, 8$:
\begin{eqnarray}
\label{order}
&&M_1 \, = \, \, M_{w=0}(\infty)(\Omega,\, \Omega), \qquad 
M_2 \, = \, \, M_{w=0}(1)(\Omega,\, \Omega), \qquad   \\
&&M_3 \, = \, \, M_{w=0}(1/4)(\Omega,\, \Omega), \qquad 
M_4 \, = \, \, M_{w=0}(w_1)(\Omega,\, \Omega), \qquad \nonumber \\
&&
M_5 \, = \, \, M_{w=0}(-1/2)(\Omega,\, \Omega), \quad  
M_6 \, = \, \, M_{w=0}(-1/4)(\Omega,\, \Omega), \qquad \nonumber \\
&&M_7 \, = \, \, M_{w=0}(0)(\Omega,\, \Omega), \qquad  
M_8 \, = \, \, M_{w=0}(w_2)(\Omega,\, \Omega) \nonumber
\end{eqnarray}

The matrices $\, M_2$, $\, M_4$, $\, M_5$, $\, M_8$,
and respectively the matrices $\, M_1$ and $\, M_7$, share the same
Jordan block form.
The Jordan block forms for $\, M_3 $ and $\, M_6 $
read respectively:
\begin{eqnarray}
\left[ \begin {array}{cccccc} 
-1&0&0&0&0&0\\
\noalign{\medskip}0&1&1&0&0&0\\
\noalign{\medskip}0&0&1&1&0&0\\
\noalign{\medskip}0&0&0&1&0&0\\
\noalign{\medskip}0&0&0&0&1&1\\
\noalign{\medskip}0&0&0&0&0&1
\end {array} \right], \, \qquad 
\left[ \begin {array}{cccccc}
 -1&0&0&0&0&0\\
\noalign{\medskip}0&1&1&0&0&0\\
\noalign{\medskip}0&0&1&1&0&0\\
\noalign{\medskip}0&0&0&1&0&0\\
\noalign{\medskip}0&0&0&0&1&0\\
\noalign{\medskip}0&0&0&0&0&1
\end {array} \right] \nonumber 
\end{eqnarray}

These matrices $\, M_i$ are the generators of a $\, 6 \times 6$ matrix representation 
of the differential Galois group of the Fuchsian differential
equation corresponding to $L_6$.
Any element of the differential Galois group is of the form:
\begin{eqnarray}
\label{g}
 M_{P(1)}^{n_1} \cdot  M_{P(2)}^{n_2} \cdot  M_{P(3)}^{n_3} 
\cdot  M_{P(4)}^{n_4}  \cdot  M_{P(5)}^{n_5} 
  \cdot  M_{P(6)}^{n_6}   \cdot  M_{P(7)}^{n_7} \cdot  M_{P(8)}^{n_8} 
\end{eqnarray}
where $\, P$ denotes an arbitrary permutation of eight elements
 and where the $\, n_i$'s are positive or negative integers. 
This looks, at first sight, like an infinite discrete group, but the closure
of this infinite set of matrices can be quite large continuous groups
like semi-direct products of $\, SL(2,\mathcal{C})$ with $\, SL(3,\mathcal{C})$, ...

Our ``global'' (800 digits, 1500 terms) calculations yield
quite involved exact connection matrices. With such large and
involved computer calculations there is always 
a risk of a subtle mistake or misprint.
At this stage, and in order to be ``even more confident''
in our results, let us recall that the monodromy matrices must satisfy one
matrix relation which will be an {\em extremely severe non-trivial check}
on the validity of these
eight matrices $\, M_i$, or more precisely their $\, (\alpha, \, \Omega)$
extensions. Actually it is known (see for instance
 Proposition 2.1.5 in~\cite{Alexa}), that 
the monodromy group\footnote[3]{Which identifies in our Fuchsian case
to the differential Galois group.}
of a linear differential equation (with $r$ regular singular points)
 is generated by a set of matrices
$\, \gamma_1, \,\gamma_2, \, \cdots, \, \gamma_r$ that satisfy 
$\, \gamma_1 \cdot \gamma_2  \cdots \gamma_r\, = \, {\bf Id}$,
where $\, {\bf Id}$ denotes  the identity matrix.
The constraint that ``some'' product of all these matrices
should be equal to the identity matrix, looks quite simple, but is, in fact,
``undermined'' by subtleties of complex analysis
on how connection matrices between non neighboring singular points
should be computed.
The fact that the prescription (\ref{neighbors},\ref{neighbors-w2})
 has given no contradictory
results on the $\tilde{\chi}^{(3)}$  singular behavior may
be an argument that our $\, M_i$'s are not ``too far'' from these
``elementary'' $\, \gamma_i$'s.
In other words, one of the products (\ref{g}) must be equal to
the identity matrix for some set of $\, n_i$'s and for some permutation $\, P$.
With the particular choice (\ref{order})  of ordering of the eight
singularities, this product, actually reads:
\begin{eqnarray}
\label{relid}
 M_1 \cdot  M_2 \cdot  M_3 
\cdot  M_4 \cdot  M_5 
  \cdot  M_6   \cdot  M_7 \cdot  M_8 \, = \, \,\, {\bf Id} 
\end{eqnarray}
Of course, from this relation, one also has seven other relations
deduced by cyclic permutations.
It is important to note that these relations (\ref{relid}) 
{\em are not verified}
by extensions
like (\ref{MalphaOmega1}), (\ref{globsq}) depending on 
two independant parameters $\, \alpha$ and $\, \Omega$, 
of the monodromy matrices $\, M_i$. If one imposes relations
(\ref{relid}) for the $(\alpha, \, \Omega)$ extensions of the $\, M_i$'s, one 
will find that, necessarily, $\, \alpha$ has to be equal
 to $\, \Omega$, but (of course\footnote[4]{A matrix identity like (\ref{relid}) yields a set of 
polynomial (with integer coefficients) relations on 
$\, \Omega\, = \, \, 2 \, i \, \pi$. The number $\, \pi$ being transcendental
it is not a solution of a polynomial with integer coefficients. These 
polynomial relations have, thus, to be {\em polynomial identities valid for any} $\, \Omega$.})
one will find that these matrix identities {\em are verified for any value of}
 $\, \Omega$, not necessarily equal
to $\, 2 \, i \, \pi$.

\subsection{Comments}
\label{comme}
The entries of the connection matrices 
have been seen to be expressed as various polynomials, or algebraic
combinations of power of $\pi$, $\, \ln(2)$, $\, \ln(N)$ ($N$ integer), 
algebraic numbers, etc.,  and more ``involved''  transcendental numbers like
(\ref{I3p}).
On the other hand, the monodromy matrices $\, M_{w=0}(w_s)$,
expressed in the same $\, (w=0)$ well-suited bases, have
entries which are rational expressions with integer coefficients  
of  $\, \alpha$ and $\, \Omega$.
To get some hint as to how this occurs, let us consider, for instance, the regular singular
point $w=1$. The local monodromy matrix is almost the unity matrix
(only one solution with log) with elements:
\begin{eqnarray}
 {\it l}(1)_{ij} \, = \, \, \, \delta_{ij} \, \,+ \Omega \cdot  \delta_{i3} \,\, \delta_{j2}
\end{eqnarray}
The product (\ref{de0a1}) giving the global monodromy matrix
will be given by
\begin{eqnarray}
\label{glob-ok}
 M_{w=0}(1)_{ij} \, = \, \, \, \,\delta_{ij}  \,\,+ \Omega \cdot C(0,1)_{i3} \cdot C^{-1}(0,1)_{2j}
\end{eqnarray}
where one can see that only the third column of $C(0,1)$ and the
second row of its
inverse will contribute. These entries have been ``recognized'' (see Appendix D).

Let us assume that there is another solution with a log term (this is not so, see
Table 1). An entry (for instance ${\it l}(1)_{65}$) of the local
monodromy matrix changes from zero to $\Omega$. In this case
equation (\ref{glob-ok}) becomes:
\begin{eqnarray}
\label{glob-assum}
 M_{w=0}(1)_{ij}\, = \, \delta_{ij}\,
 + \Omega \cdot C(0,1)_{i3} \, C^{-1}(0,1)_{2j}
\, + \Omega \cdot  C(0,1)_{i6} \, C^{-1}(0,1)_{5j} \nonumber
\end{eqnarray}
The entries $C(0,1)_{i6}$ and $C^{-1}(0,1)_{5j}$ 
will appear in the global monodromy matrix.
In fact, changing the entry ${\it l}(1)_{65}$ from zero
to $\Omega$ means that a formal solution will exhibit log's, and this
will correspond to the entries $C(0,1)_{i6}$. As a pratical rule,
we found that such entries (corresponding to solutions with log's)
can be easily ``recognized'' in contrast with the entries corresponding
to Frobenius series which will be canceled by the zero entries of
${\it l}(1)$. The entries corresponding to Frobenius series are
probably valuations of holonomic functions.

Let us now assume (for the actual situation) that the whole column
$C(0,1)_{i3}$ has unknown entries.
Recalling the fact that the product of the monodromy matrices, expressed 
in the same basis, should be equal to the identity matrix~\cite{Alexa}
(this is what we found for our eight matrices $M_i$, see (\ref{relid})),
one then expects the ``not yet guessed constants'' (i.e., the column
$C(0,1)_{i3}$) to be given by a non linear system of equations.
This is indeed what occurs for this example, and we recover that way the entries
given for this case in Appendix D.

A last remark is the following.
Right now, we have considered all the matrices (connection and therefore
monodromy matrices expressed in a unique basis) with respect to the ($w=0$) well-suited basis
of solutions. This is motivated by the physical solution $\tilde{\chi}^{(3)}$
which is known as series around $w=0$.
In fact, we can switch to another $w=\tilde{w}$ well-suited
 basis of solutions. This amounts to considering the connection
$C(\tilde{w}, w_s)=C^{-1}(0, \tilde{w})\cdot C(0,w_s)$.
For instance,  we have actually
performed the same calculations for the  $(w=1/4)$ basis of series solutions.
We have calculated all the connection matrices from  the  $(w=1/4)$ basis
to the other singular point basis series solutions, and deduced the exact
expressions of the corresponding  monodromy matrices now expressed {\em in the same} 
 $(w=1/4)$ {\em basis of series solutions}. 
It is worth noting that we get, this time, for the monodromy
$\, M_{w=1/4}(w_s)$
around singular point $\, w_s$ and expressed in the  $(w=1/4)$
basis, a matrix {\em whose entries depend rationally on}
$\, \alpha$, $\, \Omega$, {\em but, this
 time, also (except for the monodromy matrix at $w=1$) on the
 "ferromagnetic constant"} $\, I_3^{+}$.
One verifies that the product of these monodromy
matrices in the {\em same order} as (\ref{relid}),  is actually equal to the identity matrix 
when $\, \alpha \, = \, \Omega$, the matrix identity being valid for any 
value of $\, \alpha \, = \, \Omega$ (equal or not to $\, 2\, i\, \pi$),
and for any value of $\, I_3^{+}$
(equal or not to its actual value given in (\ref{I3p})).

We have similar results for the monodromy matrices 
around singular point $\, w_s$, expressed in the  $(w=\infty)$ basis, but,
now, the  monodromy matrices $\, M_{w=\infty}(w_s)$ depend on
$\, \alpha$, $\, \Omega$, and, this time, on the (not yet recognized)
constants $\, y_{41}$ and $\, x_{42}$.
Again, the product of these monodromy
matrices in the same order as (\ref{relid}),  is actually equal to the identity matrix 
when $\, \alpha \, = \, \Omega$, the matricial identity being valid for any 
value of $\, \alpha \, = \, \Omega$ (equal or not to $\, 2\, i\, \pi$)
and for any values of $\, y_{41}$ and $\, \,x_{42}$
(equal, or not, to their actual  values given in Appendix C).

\section{Mutatis mutandis: Connection matrices and singular behavior for
$\tilde{\chi}^{(4)}$}
\label{muta}
\subsection{Connection matrices}
The Fuchsian differential equation for\footnote[2]{$\tilde{\chi}^{(n)}$
is defined as $\chi^{(n)}=(1-s^{-4})^{1/4} \cdot \tilde{\chi}^{(n)}$, for $n$ even. }
$\tilde{\chi}^{(4)}$, the four-particle contribution to the
susceptibility, is given in \cite{ze-bo-ha-ma-05b}. The order ten differential
operator ${\cal L}_{10}$ associated with this differential equation
has 36 (equivalent up to isomorphisms) factorizations
(see Appendix F in~\cite{ze-bo-ha-ma-05b}). Consider, for instance,
two of these factorizations: 
\begin{eqnarray}
\label{factochi4bis}
{\cal L}_{10} \, &=& \, \, N_8 \cdot M_2 \cdot L_{25} 
\cdot L_{12} \cdot L_{3} \cdot L_0  \\
\, &=& \, \,  M_1 \cdot L_{24} 
\cdot L_{13} \cdot L_{17} \cdot L_{11} \cdot N_0 \nonumber 
\end{eqnarray}
The notations are the same as those in \cite{ze-bo-ha-ma-05b}, the $M$ operators are
of order four, the $N$ and $L$ operators are respectively of order two and one.
The two factorizations above  mean that $\, {\cal L}_{10} $ is a {\em direct 
sum} of an order eight differential operator,
 $\, {\cal L}_{8}$  $\, = \,  M_2 \cdot L_{25} 
\cdot L_{12} \cdot L_{3} \cdot L_{0}$ 
and of the order two differential operator $\, N_0$ 
(which, see~\cite{ze-bo-ha-ma-05b}, has remarkably
$\tilde{\chi}^{(2)}$ as solution):
\begin{eqnarray}
\label{factochi4ter}
{\cal L}_{10} \,  = \,\, \,\, {\cal L}_{8} \oplus  N_0
\end{eqnarray}
As was the case for $\tilde{\chi}^{(3)}$, it is thus sufficient to consider the differential
operator ${\cal L}_{8}$ for which 
a general form of $\, 8 \times 8$
 matrices, representing
$\, {\cal G}al({\cal L}_{8})$,  
the differential Galois group of ${\cal L}_{8}$, is deduced:
\begin{eqnarray}
\label{galL10}
\left[ \begin {array}{cc}
\noalign{\medskip} {\bf L} & {\bf 0}\\
\noalign{\medskip} {\bf G} & {\bf M}
\end {array} \right]
\nonumber
\end{eqnarray}
${\bf G}$, ${\bf M}$ and ${\bf L}$ are $4 \times 4$ matrices,
the latter being lower triangular. Recall that
${\cal L}_{8}$ has four known global solutions (see \cite{ze-bo-ha-ma-05b} and below).

Similarly to the calculation on $\tilde{\chi}^{(3)}$, we can, for instance, calculate 
connection matrices associated
with the correspondence between the series near $\, x \, = 16 \, w^2\, \, = \, 0$
(high temperature)
with the series near $\, x \, = 16 \, w^2\, \, = \, 1$
(ferromagnetic and antiferromagnetic critical point), and  find 
how the ``physical solution'' $\tilde{\chi}^{(4)}$ can be decomposed 
on the various well-suited bases around each 
singular point (physical or non-physical)
of the order ten Fuchsian differential equation. 

We use the factorization (\ref{factochi4bis})
to construct the basis of solutions, sequentially, as 
the four solutions corresponding to the differential operator
$L_{25} \cdot L_{12} \cdot L_{3} \cdot L_{0}$
that we call respectively $S_1$, $S_2$, $S_3$ and $S_4$.
To these solutions, we add the
four solutions coming from ${\cal L}_8$ and inherited from the differential operator $\, M_2$,
 that we call $S_5$, $\cdots$, $S_8$.
Here, again, an optimal
choice of basis is made in order to have as many zeroes as possible in
the connection matrix with as "simple" entries as possible.
The basis of solutions at $x=0$ and $x=1$ (with respectively $t=x$ and
$t=1-x$) have similar forms and read:
\begin{eqnarray}
S_1 (t) &=& \, \,  1, \qquad  \qquad 
S_2 (t) = \, \,  {\rm eq.}(33)\,\, {\rm in}\,\, \cite{ze-bo-ha-ma-05b},    \nonumber \\
S_3 (t) &=& \, \,  {\rm eq.}(32)\,\, {\rm in}\,\, \cite{ze-bo-ha-ma-05b}, \qquad  \qquad 
S_4 (t) =\, \,   {\rm eq.}(43)\,\, {\rm in}\,\, \cite{ze-bo-ha-ma-05b},          \nonumber \\
S_5 (t) &=& \, \,  {\rm see \,\, below}, \qquad  \qquad 
S_6 (t) =\, \,   S_5 \,(\ln(t/16) +a_1)\, +S_{60} \nonumber \\
S_7 (t) &=&\, \,    \,\left(\ln(t/16)^2\,+2a_1\ln(t/16)\,+a_2 \right) \cdot  S_5   \nonumber \\
&&\quad  +2S_{60} \left(\ln(t/16)+a_1 \right)   +S_{70}  \nonumber \\
S_8 (t) &=&\, \,   \,\left( \ln(t/16)^3   \,
 +3 \, a_1\ln(t/16)^2+3a_2\ln(t/16)\,+a_3 \right) \cdot  S_5 \nonumber \\
&& \quad  +3\,  \left(\ln(t/16)^2\, +2a_1\ln(t/16)\, +a_2 \right) \cdot S_{60} \nonumber \\
&& \quad  +3\,  \left(\ln(t/16)\, +a_1 \right)\cdot S_{70}  + S_{80}  \nonumber 
\end{eqnarray}
where the constants $a_1$, $a_2$ and $a_3$ and the series read, near $x=0$
\begin{eqnarray}
a_1 &=& 79/60, \quad \quad \quad a_2=-751/1800, \quad \quad \quad a_3=-10619/375, \nonumber \\
S_5^{(0)} (t) &=&  \, \,  [0, 0, 1, 45/32, 425/256, 945/512, \cdots ],           \nonumber \\
S_{60}^{(0)} (t) &=&  \, \,  [0, 2/3, 0, 2353/13440, 121619/322560, \cdots ],     \nonumber \\
S_{70}^{(0)} (t) &=&  \, \,  [8, -119/45, 0, -560333/1411200,  \cdots ],     \nonumber \\
S_{80}^{(0)} (t) &=&  \, \,  [0, 0, 0, 0, -127639044817/85349376000, \cdots ]     \nonumber 
\end{eqnarray}
and, near $x=1$ :
\begin{eqnarray}
a_1 &=& \, \,  35/6, \quad \quad a_2=107/9, \quad \quad a_3=-1051745657/749700 \nonumber \\
S_5^{(1)} (t) &=&  \, \,  [1, -1/4, -7/64, -45/256, -3385/16384, \cdots ],           \nonumber \\
S_{60}^{(1)} (t) &=& \, \,   [0, 7/120, -3809/13440, 42401/16120, 9271027/18923520, \cdots ],     \nonumber \\
S_{70}^{(1)} (t) &=& \, \,   [0, 1099/75, 741847/78400, 218499331/101606400,  \cdots ],     \nonumber \\
S_{80}^{(1)} (t) &=& \, \,   [0, 0, 0, -37462660457/592220160, \cdots ]     \nonumber 
\end{eqnarray}
The connection matrix between $x=0$ and $x=1$ comes out as:
\begin{eqnarray}
C(0, 1) \, = \, \, 
\left [\begin {array}{cc} 
{\bf 1} & {\bf 0} \\
\noalign{\medskip}
 {\bf A} & {\bf B} 
\end {array}\right]
\end{eqnarray}
where $\, {\bf 1} $ denotes the $\, 4 \times 4$ identity matrix
and $\, {\bf 0}$ denotes the $\, 4 \times 4$ zero matrix.
The $\, 4 \times 4$ identity matrix corresponds 
to the fact the four solutions $S_1$, $\cdots$, $S_4$
are {\em global solutions}.
The two lower $4 \times 4$ blocks read:
\begin{eqnarray}
{\bf A}=\left [\begin {array}{cccc} 
a_{51} & a_{52} & -{{5}\over{2}} & a_{54} \\
\noalign{\medskip}
0 & {{2}\over{3}}\pi  & 0 & {{1}\over{32}}  \\
\noalign{\medskip}
a_{71} & 0 & a_{73} & 0 \\
\noalign{\medskip}
a_{81} & -\pi^3 & a_{83} & a_{84} \\
\end {array}\right],\,\, 
 {\bf B}=
\left [\begin {array}{cccc}
0 & 0 & 0 & -{{1}\over{2\pi^3}} \\
\noalign{\medskip}
0 & 0 & -{{1}\over{2\pi}} & 0  \\
\noalign{\medskip}
0 & -{{\pi}\over{2}} & 0 & 0 \\
\noalign{\medskip}
-{{\pi^3}\over{2}} & 0 & 0 & 0 \\
\end {array}\right] \nonumber
\end{eqnarray}
with
\begin{eqnarray}
a_{71} = {\frac{\pi^2}{6}}-{\frac{2422}{225}}, \quad
a_{73} = {\frac{5\pi^2}{6}}+{\frac{2422}{225}}, \quad
a_{84} = -{\frac{\pi^2}{32}}-{\frac{1211}{600}} \nonumber
\end{eqnarray}
The ``not yet recognized'' entries of this matrix read:
\begin{eqnarray}
a_{51} &\simeq & -17.882936774520, \quad a_{52} \simeq 7.767669067696, \nonumber \\
a_{54} &\simeq& 0.530951641617, \quad a_{81} \simeq -92.773462923758, \nonumber \\
a_{83} &\simeq& 77.887072991056 \nonumber
\end{eqnarray}

Here again, the block structure of the connection matrix relies
on the factorization of ${\cal L}_8$ and on the ``sequential'' building of the
solutions. The block matrix ${\bf B}$ represents, specifically, the connection between
the solutions inherited from $\, M_2$ at both points $x=0$ and $x=1$. This fourth order differential
operator $\, M_2$ in ${\cal L}_8$ (corresponding to $\tilde{\chi}^{(4)}$) is structurally very
 similar (see the remark at end of Appendix B)
to operator $\, Y_3$ in $L_6$ ($\tilde{\chi}^{(3)}$).
Similarly to $\tilde{\chi}^{(3)}$ case, a ferromagnetic (and anti-ferromagnetic) 
constant (see (\ref{constchi4}) below) is
localized at the fifth line.

We have also computed the connection matrices\footnote[3]{The matching
points are taken in the lower half-plane of the variable $x$.}
(not given here) between
the solutions at $x=0$ and respectively $x=4$ (corresponding to
Nickel's non-physical singularities) and $x=\infty$ (corresponding to
the non-physical singularities $s=\pm i$).
Denoting by $M_{x=0}(0)$, $M_{x=0}(1)$, $M_{x=0}(4)$ and $M_{x=0}(\infty)$,
the monodromy matrices expressed in the same $x=0$ well-suited basis
obtained with similar conjugation like (\ref{de0a1}), one obtains:
\begin{eqnarray}
M_{x=0}(\infty) \cdot M_{x=0}(4) \cdot M_{x=0}(1) \cdot M_{x=0}(0) \, =\, {\bf Id}
\end{eqnarray}
This identity is valid irrespective of the still unknown
constants.

\subsection{Singular behavior of $\tilde{\chi}^{(4)}$}
The particular physical solution corresponding to
$\tilde{\chi}^{(4)}=\tilde{\chi}^{(2)}/3 + \Phi_8$
(see \cite{ze-bo-ha-ma-05b}) is given,
in terms of the basis chosen at the point $x=0$, by:
\begin{eqnarray}
\Phi_8 \, =\,\, \, {{1}\over{384}} \cdot \Bigl(    
5\, S_1^{(0)} -5\, S_3^{(0)} -2\,S_5^{(0)} \Bigr)
\end{eqnarray}
At the ferromagnetic, and anti-ferromagnetic, critical point  $x=1$, the
solution can be deduced from the above connection matrix and reads:
\begin{eqnarray}
\Phi_8 \, =\,
-{\frac {1}{384}}\,\left (2\, a_{51}-5\right )\cdot  S_1^{(1)}\, 
-{\frac {a_{52}}{192}} \cdot  S_2^{(1)}\, 
-{\frac {a_{54}}{192}} \cdot  S_4^{(1)}\, 
+{\frac {1}{384 \pi^3}} \cdot  S_8^{(1)} \nonumber
\end{eqnarray}
Here again, the above decomposition corresponds to an expansion at the point $x=1$
of the triple integral defining $\tilde{\chi}^{(4)}$.

From this solution, the singular part of $\tilde{\chi}^{(4)}$
reads (with $t=1-x$):
\begin{eqnarray}
\label{singuchi4}
\tilde{\chi}^{(4)}({\rm singular}, 1)  \,  & = &  \,\,\,\, \,{{ I_4^{-} } \over {t}}
\,\, \, + {{ 1} \over {384\, \pi^3}}\, S_5^{(1)} \cdot \ln^3(t)\, \\
&&   - {{ 1} \over {32\, \pi^3}}\,
\Bigl((\ln(2) -{{35}\over{24}}) S_5^{(1)} 
-{{35}\over{24}} S_{60}^{(1)} \Bigr) \cdot    \ln^2(t) \nonumber \\
&&     + {{ 1} \over {8\, \pi^3}}\cdot
\Bigl(( \ln(2)^2- {{35}\over{12}} \ln(2) +{{107}\over{144}}) \cdot  S_{5}^{(1)} \nonumber \\
&& \quad \quad -({{1} \over {2}} \ln(2)-{{35} \over {48}})\cdot S_{60}^{(1)}
+{{1} \over {16}} S_{70}^{(1)} \Bigr) \cdot
 \ln(t) \nonumber \\
 &&   + {{ 1} \over {48\, \pi}}\,
 _{2}F_1 \left( 1/2, -1/2; 2; t \right) \cdot \ln(t) \nonumber
\end{eqnarray}
The constant \cite{wu-mc-tr-ba-76} $I_4^{-}$ reads,
in terms of the ``not yet recognized'' numbers $\, a_{52}$,
$a_{54}$:
\begin{eqnarray}
\label{constchi4}
I_4^{-} \, = \, \, \, \, {{1}\over{36\pi}} +{{a_{52}}\over{128}}\,\,
- {{a_{54}\,\pi}\over{24}}\,  \,\, \simeq  \,\, \, \, 0.0000254485110658 \,\, \cdots
\end{eqnarray}
The first term at the right-hand-side of (\ref{constchi4}) comes
from $\tilde{\chi}^{(2)}$, as 
well as the last term in (\ref{singuchi4}).

Similarly, the singular behavior
of the physical solution $\tilde{\chi}^{(4)}$
at the other singular points can easily be obtained from the corresponding
connection matrices (not given here).
At the singular point $x=4$, the
physical solution behaves like (with $t=4-x$):
\begin{eqnarray}
\tilde{\chi}^{(4)}({\rm singular}, 4)  \,  & = &  \,
\,-{\frac{i \cdot t^{13/2}}{2^{10}\cdot 3^2\cdot 5005}}  \,
\Bigl(1+{{5}\over{4}} t+{{261}\over{272}} t^2+ \cdots    \Bigr)
\end{eqnarray}
confirming the calculations in \cite{nickel-00}.

The singular behavior of $\tilde{\chi}^{(4)}$ at the singular
point $x=\infty$
reads (with $t=1/x$):
\begin{eqnarray}
\label{singchi4inf}
&& \tilde{\chi}^{(4)}({\rm singular}, \infty)  \,   =   \,\,-20\, i \,\cdot t^{-1/2} \cdot  
 \Bigl(  A_0 + 3  A_1  \cdot \ln(t)   \\
&&  +3 \Bigl((a_1-4\ln(2))\cdot S^{\infty}_5 \, +S^{\infty}_{60} \Bigr)\cdot \ln^2(t) 
  + S^{\infty}_5 \cdot \ln^3(t)   \Bigr) \nonumber \\
&& + {\frac{ (-t)^{-1/2}}{36 \pi}} \Bigl( 1
+{\frac{3\,t}{4}}\, 
_{2}F_1 \left( 1/2, 5/2; 2; t \right) \cdot \ln(-t) 
 -{\frac{9\pi \,t}{16}}\,\sum_{n=0}^\infty b_n\, t^n 
\Bigr) \nonumber
\end{eqnarray}
with
\begin{eqnarray}
A_1 &=& {{2}\over{5}}(2K-1)\cdot  S^{\infty}_{41}
 +  \Bigl( 16\ln^2(2)-8a_1\ln(2)+a_2 \Bigr)\cdot S^{\infty}_5  \nonumber \\
&+& 2(a_1-4\ln(2))\cdot S^{\infty}_{60} +3S^{\infty}_{70} \nonumber \\
A_0 &=&
2\pi^3 \Bigl( i_{52}+i\, {{24}\over{\pi^2}}(2K-1) \Bigr) S^{\infty}_2 /5 \nonumber \\
&-& \Bigl( -{{48}\over{\pi^2}}(2K-1) +i\, (5+2r_{53}) \Bigr)\pi^3 S^{\infty}_{3}/5 \nonumber \\
&-& \Bigl( 64\ln^3(2)-48a_1\ln^2(2)+12a_2\ln(2)-a_3 \Bigr) \cdot S^{\infty}_5 \nonumber \\
&+& {\frac{6}{5}}(2K-1)\cdot S^{\infty}_{40}+3\Bigl(16\ln^2(2)-8a_1\ln(2)+a_2\Bigr)\cdot S^{\infty}_{60} \nonumber \\
&+& 3(a_1-4\ln(2))\cdot S^{\infty}_{70} + S^{\infty}_{80} \nonumber
\end{eqnarray}
\begin{eqnarray}
b_n= {\frac{\Gamma(n+1/2) \Gamma(n+5/2)}{\Gamma(n+2) \Gamma(n+1)}} 
\Bigl(\Psi(n+2)+\Psi(n+1)-\Psi(n+{{5}\over{2}})-\Psi(n+{{1}\over{2}}) \Bigr) \nonumber
\end{eqnarray}
where $K=0.915965\cdots$ is {\em Catalan's constant} and the other parameters,
constants and series are: 
$a_1=2/5-\pi \, i$, $a_2=1-\pi^2-4\pi\,i/5$, $a_3=-6\pi^2/5+48193/7500+
\pi(\pi^2-3)\,i$, $i_{52}=-0.740250494 \cdots$,
$r_{53}=2.225246651 \cdots$, and
\begin{eqnarray}
S^{\infty}_2 &=& {\frac{1-6t+2t^2}{2(t-1)}}, \quad
 \quad S^{\infty}_3 = {\frac{3-12t+8t^2}{8(t-1)^{3/2}}} \nonumber \\
S^{\infty}_{40} &=& [2, 41/2, 313/48, 3047/480, \cdots ], \nonumber \\
S^{\infty}_{41} &=& [1, -25/2, -61/8, -129/16, \cdots ], \nonumber \\
S^{\infty}_{5} &=& [0, 1, 7/10, 47/64, 981/1280, \cdots ], \nonumber \\
S^{\infty}_{60} &=& [0, 0, 161/300, 2039/4800, \cdots ], \nonumber \\
S^{\infty}_{70} &=& [0, 0, 1847/18000, 2627/36000, \cdots ], \nonumber \\
S^{\infty}_{80} &=& [0, 0, 0, 14423879/7200000, \cdots ] \nonumber 
\end{eqnarray}
The last bracket in (\ref{singchi4inf}) comes from $\tilde{\chi}^{(2)}$.

Having the singular part of $\tilde{\chi}^{(4)}$ at the ferromagnetic and
anti-ferromagnetic critical points, it is straightforward to obtain
the asymptotic behavior of the series coefficients.
This time, one needs the form of the coefficients in the expansion
of $\ln^3(1-x)$ that we find to be\footnote[4]{An asymptotic form can
be obtained using various packages available at http://algol.inria.fr/libraries/software.html
like the command ``equivalent'' in gfun~\cite{gfun}, see details
in \cite{salvy-91,salvy-91b}.}
\begin{eqnarray}
\ln^3(1-x)\, = \,\, \, \, \sum_{n=3}^{\infty} \Bigl(
-{{3}\over{n}} \, \Bigl( \Psi(n)+\gamma \Bigr)^2+
{{\pi^2}\over{2n}}-{{3}\over{n}}\, \Psi(1,n) \Bigr)\cdot x^n
\end{eqnarray}
where $\Psi(1,n)$ is the first derivative of $\Psi(n)$. Expanding $\Psi(n)$
and $\Psi(1,n)$ up to $1/n^2$ for large values of $n$, one obtains the
following asymptotic behavior for the coefficients of
 the $\tilde{\chi}^{(4)}$ series: 
\begin{eqnarray}
&&c(n)  \, \simeq  \, \, I_4^{-} \, 
-{{\ln^2(n)} \over{ 128 \pi^3\,n}} 
 +{{\ln(n)} \over{ 128 \pi^3\,n^2}} \nonumber \\
&&\qquad \qquad \quad  -{{b_1\,\ln(n)} \over{ 64 \pi^3\,n}} 
 -{{b_2} \over{ 2304 \pi^3\,n}}  
  +{{b_1-1} \over{ 128 \pi^3\,n^2}}  +\cdots \nonumber
\end{eqnarray}
where:
\begin{eqnarray}
b_1 &=&\, \,\,  \gamma + 4\ln(2) -{{35}\over{6}},  \nonumber \\
b_2 &=&\, \, \, 288 \ln^2(2)+144 \gamma \ln(2)
+18\gamma^2 -210 \gamma
-840 \ln(2) +45\pi^2\, +214 \nonumber
\end{eqnarray}

\section{$\tilde{\chi}^{(1)}+\tilde{\chi}^{(3)}$
versus $\tilde{\chi}$ at scaling}
Thus far we have discussed, in Sections 4 and 6.2 the mathematical aspects of
the solutions to the Fuchsian differential equations for $\tilde{\chi}^{(3)}$
and $\tilde{\chi}^{(4)}$.
However, the physics implications of the solutions we have obtained
call for some remarks near the physical critical points.
Taking, as an example, the ferromagnetic singularity for $\tilde{\chi}^{(3)}$,
the sum of the first two $n$-particle terms behave at $\tau \simeq 0$ as:
\begin{eqnarray}
\label{chi1chi3tau}
\tilde{\chi}^{(1)}+\tilde{\chi}^{(3)} \,& \simeq & \,\,  
{{1+I_3^{+}} \over {  \tau^2}}\,  \,
- {{ \ln^2(\tau)} \over {16\, \pi^2}}\, \,
 +  \left( \ln(2) \,-{{23}\over{24}} \right)\cdot  {\frac{\ln(\tau)}{4\,\pi^2}}  \\
&&
+{\frac{11}{48}}+{\frac{3}{8}}I_3^{+}-{\frac{1}{4\pi^2}}
\left( \ln^2(2)-{{23}\over{12}}\ln(2)+{{14}\over{144}}  \right) +\cdots \nonumber 
\end{eqnarray}
The exact susceptibility, as reported in [16], yields for the
normalized susceptibility $\, \tilde{\chi} $:
\begin{eqnarray}
\label{chitau}
&&\tilde{\chi} \, =\,\,\,  {{s} \over {(1-s^4)^{1/4}}} \cdot \chi \,\,  =\,\,\, 
 {\frac{ \left(\tau+\sqrt{1+\tau^2}\right)^{-1/2}}
{(1+\tau^2)^{1/8}}} \times  \\
&& \quad \quad \quad \Bigl(
c_1\, \tau^{-2} \,F_{+}(\tau) \, +{\frac{\tau^{-1/4}}{\sqrt{2}}}\,
\sum_{p=0}^{\infty}\sum_{q=p^2}^{\infty}\, b_{+}^{(p,q)} \cdot  \tau^q\, \ln^p(\tau) \Bigr)
\nonumber
\end{eqnarray}
where $c_1=1.000815260 \cdots $ is given with some 50 digits in \cite{or-ni-gu-pe-01b}.
$F_{+}(\tau)$ and $b_{+}^{(p,q)}$ are given in \cite{or-ni-gu-pe-01b}.
The constants $1+I^{+}_3$ and $c_1$ verify $1+I^{+}_3+I^{+}_5=c_1$ with 9
digits, $I^{+}_5$, corresponding to $\chi^{(5)}$, is the constant given 
in \cite{wu-mc-tr-ba-76} (and with some 30 digits
in \cite{nickel-99}).
Thus, and as suggested in \cite{wu-mc-tr-ba-76}, the partial sums of the $\chi^{(n)}$ would
converge rapidly  to the full $\chi$.
Furthermore, adding $\chi^{(3)}$ term has resulted in a series expansion
that reproduces the first 24 terms of $\chi$ to be compared with only eight first terms for
$\chi^{(1)}$ series.

However, equation (\ref{chitau}) shows a $\tau^{-1/4}$ divergence as an overall factor
to the logarithmic singularities. This structure, absent in (\ref{chi1chi3tau}),
could suggest, in the most pessimistic scenario, 
 that the $n$-particle sequence is perhaps useless in understanding 
scaling corrections and that one should be cautious in accepting
the conclusions of studies of higher field derivatives of the
susceptibility, based on similar $n$-particle
representations~\cite{co-tr-wu-77,abraham-77}.
The same situation occurs for the low temperature regime when we compare
the first two $n$-particle terms ($\tilde{\chi}^{(2)}$ and $\tilde{\chi}^{(4)}$)
with the full $\tilde{\chi}$ at scaling \footnote{For the leading amplitude,
$\tilde{\chi}^{(2)}$ and $\tilde{\chi}^{(4)}$ give
$1/12\pi + I_4^{-} \simeq 1.0009593\cdots /12\pi$ which is very close to
$1.0009603\cdots /12\pi$ for the full $\tilde{\chi}$ \cite{nickel-99}. }.

This observation raises several profound issues, which we do not address here.
One is how the logarithmic terms in the entire sum add up to make the
$\tau^{-1/4}$ divergence be factored out.
If one assumes that the other $\tilde{\chi}^{(2n+1)}$ terms share the same singularity
structure as $\tilde{\chi}^{3}$, in particular the occurrence (in variable $\tau$ or $s$)
of only {\em integer} critical exponents at the ferromagnetic critical point,
the $\tau^{-1/4}$ divergence, as an overall factor, implies the
following correspondence :
\begin{eqnarray}
\sum_{n=1}^{\infty} \sum_{m=0}^{N(n)} \alpha_{n,m} \cdot S_{n,m} (\tau) \, \ln^m(\tau)
 \quad \rightarrow \quad 
\tau^{-1/4} \cdot
\sum_{p=0}^{\infty}\sum_{q=p^2}^{\infty}\, b_{+}^{(p,q)} \, \tau^q\, \ln^p(\tau)
\nonumber
\end{eqnarray}
with $S_{n,m} (\tau)$ analytical at $\tau=0$ and  $\alpha_{n,m}$  numerical coefficients.
$N(n)$ is the maximum power of logarithmic terms occurring in the solution around 
the ferromagnetic point of the differential equation of $\tilde{\chi}^{2n+1}$.
This correspondence requires probably a {\em very particular structure}
in the successive differential equations.
Obtaining the differential equation for $\tilde{\chi}^{(5)}$ (or 
for $\tilde{\chi}^{(6)}$), and obtaining much larger series for the
full susceptibility $\, \chi$, 
will certainly help to guess such a structure and understand the susceptibility
of the two-dimensional Ising model which continues to be a
treasure-trove of profound insights into both the mathematics and physics
of integrable systems.

Let us note that the phenomenon we have discussed may be more widespread
than that observed here. If so, a whole new chapter could be opened on
field-theoretical expansions.
The challenging problem one faces here is to link {\em linear 
and non linear} descriptions of a physical problem, namely the
description in terms of an infinite number of holonomic (linear)
expressions for a physical quantity of a non linear nature.
Actually the latter is ``Painlev\'e like'' since its
series expansion can be obtained from a 
program of {\em polynomial growth} which uses
 exclusively a quadratic finite difference double recursion
generalizing the Painlev\'e equations \cite{or-ni-gu-pe-01b,or-ni-gu-pe-01}.
The difficulty to link holonomic versus 
non-linear descriptions of physical problems is typically
the kind of problems one faces with the Feynman diagram approach of particle
physics, but the susceptibility of the Ising model is, 
obviously, the simplest non trivial example to address such an
important issue.

\section{Conclusion}
\label{conclu}
We have introduced a simple and very efficient method to calculate
numerically, with an arbitrary number of digits, the connection matrices
between the independent solutions, defined at two singular points, of
differential equations of quite high orders.
We have considered the order seven, and ten, Fuchsian ODE's
corresponding to the three and four particle contribution 
to the magnetic susceptibility of the Ising model.
The entries of the connection matrix between two regular singular points
have been obtained in floating point form and most of them have been recognized,
particularly those that show up in the singular behavior of the
physical solutions.
They are expressed as polynomial, or algebraic, combinations
of $\pi$, $\ln(2)$, $\cdots$, radicals, and more involved numbers (not yet recognized)
such as the "ferromagnetic constant" (\ref{I3p}).
The method allows us to obtain the series expansions of the physical solutions
$\tilde{\chi}^{(3)}$ (and $\tilde{\chi}^{(4)}$) around any other
regular singular point, besides the already known series around
$w=0$.
We obtained, in this way, near each singular point all the dominant, and
subdominant, singular behaviors of the physical solutions.
Such subdominant singular behavior is certainly hard to obtain from
series analysis.
At the newly found quadratic singularities of the
differential equation, we showed that
the physical solution $\tilde{\chi}^{(3)}$ itself {\em is not singular}.
Also note, at $w=1/4$, that the behavior in $(1-4w)^{-3/2}$ corresponding
to the largest critical exponent for the ODE is {\em actually
absent} in the physical solution.
Note the remarkable fact that the factorization of differential operator
$L_7$ (and ${\cal L}_{10}$) associated with $\tilde{\chi}^{(3)}$ (respectively
$\tilde{\chi}^{(4)}$) shows clearly the differential operator
responsible of the non-physical singularities given in
\cite{nickel-99,nickel-00} and the newly
found quadratic numbers \cite{ze-bo-ha-ma-04}.
In both cases ($\tilde{\chi}^{(3)}$ and
$\tilde{\chi}^{(4)}$), these non-physical singularities are carried by
the differential operator $Z_2\cdot N_1$ (respectively
$L_{25}\cdot L_{12} \cdot L_3 \cdot L_0$) occurring at the right of $L_7$
(respectively ${\cal L}_{10}$).

The physical solutions $\tilde{\chi}^{(3)}$ (and $\tilde{\chi}^{(4)}$)
being known as series around $w=0$, the growth behavior of the
corresponding series coefficients should be controlled by the
singular behavior at the nearest singular points which are the ferromagnetic
and anti-ferromagnetic critical points in both cases ($w=\pm 1/4$ and $x=1$).
This growth is easily found from the expansion around the ferromagnetic
and anti-ferromagnetic points.

The connection matrices we have obtained allow us to relate the solutions
around any given singular point to a common (non-local) basis of solutions.
In this respect, we have obtained the exact expression of all
the monodromy matrices, expressed in the {\em same} basis, and we have
seen that
they are simple matrices with rational function entries.
In a forthcoming publication~\cite{Weil}, we will 
give the whole structure of the differential 
Galois group for the two previous Fuchsian differential equations.

As far as the physics implications of the solutions are concerned, we 
have compared the corrections to scaling at the ferromagnetic point given by
the first two terms ($\chi^{(1)}$ and $\chi^{(3)}$) with the full $\chi$.
Qualitative difference is found raising profound issues on the $n$-particle
representation of the susceptibility.
The same observation occurs for the antiferromagnetic point, and also for the
low temperature regime.

\vskip 0.2cm
\textbf{Acknowledgments} We thank Jacques-Arthur Weil for many valuable
comments on differential Galois group and connection matrices.
We would like to thank B. Nickel for his inspired comment
on solution $\, S_3$, many exchanges of informations and for pointing
some misprints in our series-solutions, in the earlier version of the
manuscript.
We would like to thank A. J. Guttmann, I. Jensen, and W. Orrick
for a set of useful comments on the singularity behavior 
of physical solutions. One of us (JMM) would like to thank B.M. McCoy
for many extensive discussions on the problem of the holonomic description
of non-linear problems. 
We would like to thank an anonymous referee for raising important
points on the physics implications of our results that we discussed
in Section 7.
(S. B) and (S. H) acknowledge partial support from PNR3.

\section{Note added in the Proofs}
\label{note}

After completion of the revised version of our manuscript 
we were told that, as consequence of the work of B. M. McCoy, C. A. Tracy and 
T.T. Wu, the two transcendental numbers $\, I_3^{+}$
and  $\, I_4^{-}$ can actually be written 
in terms of polylogarithms, namely 
the Clausen function $\,Cl_2$ and of the Riemann zeta
function, as follows :
\begin{eqnarray}
&&I_3^{+} \, = \, \, {{1} \over {2 \pi^2 }}\cdot \Bigl( {{\pi^2} \over {3}} \,
 +2 \, -3 \sqrt{3}\cdot  Cl_2({{\pi} \over {3}})  \Bigr),
 \, \quad  \, \, \,    Cl_2(\theta) \, = \, \, 
\sum^{\infty}_{n=1}  \, {{\sin(n\,\theta) } \over {n^2 }} \nonumber \\
&&I_4^{-} \, = \, \,{{1} \over {16 \pi^3 }}\cdot \Bigl( {{4\,\pi^2} \over {9}} \, -{{1} \over {6}} 
 -{{7} \over {2}}\cdot \zeta(3)
 \Bigr) \nonumber
\end{eqnarray}
The derivation of these results has never been published but 
these results appeared in a conference proceedings~\cite{Tracy}.
We have actually checked that $\, I_3^{+}$
and  $\, I_4^{-}$ we got from the calculations displayed in our paper
as floating numbers with respectively 421 digits and 431 digits accuracy
{\em are actually in agreement} with the previous two formula. 
These two results provide a clear answer to the question of how ``complicated
and transcendental'' some of our constants occurring in the entries of the
connection matrices can be. These extremely interesting results
are not totally surprising when one recalls the deep link between 
 zeta functions, polylogarithms and hypergeometric series~\cite{Tanguy,Tanguy2,Tanguy3,Tanguy4}.

\vskip 0.5cm

\vskip 0.5cm

\section{Appendix A}
\label{appendA}
We give, in this Appendix, the explicit expressions of the 
differential operators $X_1$ and $Z_2$ and $Y_3$.
The order one differential operator reads
\begin{eqnarray}
X_1 \, = \, \,\,   {{d} \over {dw}} \, +  {\frac{p_0}{p_1}}
\end{eqnarray}
with:
\begin{eqnarray}
p_1 & =&
 \left( -1+w \right)  \left( 4\,w-1 \right)  \left( 1+2\,w \right) 
 \left( 4\,w+1 \right)  \left( 1+3\,w+4\,{w}^{2} \right)  \nonumber \\
&& \left(1 -3\,w-18\,{w}^{2}+104\,{w}^{3}+96\,{w}^{4} \right) \nonumber \\
&& \Bigl( 1-7\,w-4\,{w}^{2}-47\,{w}^{3}+36\,{w}^{
4}+280\,{w}^{5}+160\,{w}^{6}+256\,{w}^{7} \Bigr)  \nonumber \\
p_0  &=&
w \cdot \Bigl(-58 +909\,w+3284\,{w}^{2}-24711\,{w}^{3}-72352\,{w}^{4}+181016\,{w}^{5} \nonumber \\
&& +1251768\,{w}^{6} +2852880\,{w}^{7}+1454592\,{w}^{8}-11455616\,{w}^{9} \nonumber \\
&& -31712256\,{w}^{10}-20418560\,{w}^{11}+20840448\,{w}^{12}+34963456\,{w}^{13} \nonumber \\
&& +30146560\,{w}^{14}+15728640\,{w}^{15}
  \Bigr) \nonumber
\end{eqnarray}

The order two differential operator $Z_2$ is
\begin{eqnarray}
Z_2 \, = \, \,\, \,  {\frac{1}{p_2}}\, \sum_{n=0}^2\, p_n\cdot {{d} \over {dw^n}} 
\end{eqnarray}
where the polynomials $p_i$'s, now, read:
\begin{eqnarray}
p_2 &=& \,\,  w \cdot \left(4\,w-1\right )^{2}\left (4\,w+1\right )
\left (1+3\,w+4\,{w}^{2}\right )\left (-1+w\right )
\left (1+2\,w\right ) \nonumber \\
&& \left (1-3\,w-18\,{w}^{2}+104\,{w}^{3} +96\,{w}^{4}  \right) \nonumber \\
p_1 &=& \, \, \left (4\,w-1\right )
\Bigl (1-6\,w-111\,{w}^{2} -108\,{w}^{3} +1080\,{w}^{4} -4488\,{w}^{5} \nonumber \\
&& -40368\,{w}^{6}-94272\,{w}^{7}-48384\,{w}^{8} +72704\,{w}^{9} +49152\,{w}^{10} \Bigr) \nonumber \\
p_0 &=&\, \,  4+48\,w-276\,{w}^{2}-1520\,{w}^{3}-3192\,{w}^{4}-4224\,{w}^{5}-71552\,{w}^{6} \nonumber \\
&& -307200\,{w}^{7}-239616\,{w}^{8} +98304\,{w}^{9} +98304\,{w}^{10} \nonumber
\end{eqnarray}

The order three differential operator $\, Y_3$ is given by
\begin{eqnarray}
\label{Y3}
Y_3 \, = \, \, \, \, {\frac{1}{p_3}}\, \sum_{n=0}^3\, p_n\cdot  {{d^n} \over {dw^n }}
\end{eqnarray}
where the polynomials $p_i$'s, now, read:
\begin{eqnarray}
\label{d3}
&& p_3 = \, 
{w}^{2} \cdot \left (w-1\right )\left (1+2\,w\right )
\left (1+3\,w+4\,{w}^{2}\right )  \\
&& \left (4\,w-1\right )^{3}\left (4\,w+1\right )^{3}\left (96\,{w}^{4}+
104\,{w}^{3}-18\,{w}^{2}-3\,w+1 \right )^{3} \nonumber \\
&& \Bigl( 1+19\,w-368\,{w}^{2}-3296\,{w}^{3}+17882\,{w}^{4}+272599\,{w}^{5}+160900\,{w}^{6} \nonumber \\
&& -6979208\,{w}^{7}+7550800\,{w}^{8}+203094872\,{w}^{9}-278920192\,{w}^{10}  \nonumber \\
&& -3959814304\,{w}^{11}-2115447424\,{w}^{12}+20894729472\,{w}^{13} \nonumber \\
&& +39719728128\,{w}^{14}+20516098048\,{w}^{15}+ 256763363328\,{w}^{16} \nonumber \\
&& -327065010176\,{w}^{17}-8810227761152\,{w}^{18}  +414933057536\,{w}^{19} \nonumber \\
&& +116411936538624\,{w}^{20}  +296827723186176\,{w}^{21}+317648030138368\,{w}^{22} \nonumber \\
&& +179148186189824\,{w}^{23}+194933533179904\,{w}^{24}+112931870081024\,{w}^{25} \nonumber \\
&& -55246164328448\,{w}^{26}+11063835754496\,{w}^{27}+1511828488192\,{w}^{28}   
  \Bigr) \nonumber
\end{eqnarray}
\begin{eqnarray}
&& p_2 =\,  w \cdot 
\left (4\,w-1\right )^{2}
\left (4\,w+1\right )^{2}\left (96\,{w}^{4}+104\,{w}^{3}-18\,{w}^{2}-3
\,w+1\right )^{2}  \nonumber \\
&& \Bigl( 6+102\,w-2018\,{w}^{2}-23962\,{w}^{3}+242904\,{w}^{4}+2575633\,{w}^{5} \nonumber \\
&& -12389010\,{w}^{6}-178413527\,{w}^{7}+80727412\,{w}^{8}+6252221348\,{w}^{9} \nonumber \\
&& +2456938016\,{w}^{10} -178278888104\,{w}^{11}-103902989696\,{w}^{12} \nonumber \\
&& +3814815965856\,{w}^{13}+1524977514176\,{w}^{14}-67400886678400\,{w}^{15} \nonumber \\
&& -74115827788032\,{w}^{16}+797710351468032\,{w}^{17}+2324376661856256\,{w}^{18} \nonumber \\
&& -1561280104050688\,{w}^{19}-16314064973299712\,{w}^{20} \nonumber \\
&& -27005775986622464\,{w}^{21}-40259640226480128\,{w}^{22} \nonumber \\
&& +35764751009841152\,{w}^{23}  +1007304244270727168\,{w}^{24} \nonumber \\
&& +1460771505523654656\,{w}^{25}-13359756413056843776\,{w}^{26} \nonumber \\
&& -63988213537189134336\,{w}^{27}-116684614339309600768\,{w}^{28} \nonumber \\
&& -75710498024932245504\,{w}^{29}+57121462326803824640\,{w}^{30} \nonumber \\
&& +132479693600191414272\,{w}^{31} +111232702128767107072\,{w}^{32} \nonumber \\
&& +106152703871500156928\,{w}^{33}+83508376521540632576\,{w}^{34} \nonumber \\
&& +10084606300752183296\,{w}^{35} -9404395631251816448\,{w}^{36} \nonumber \\
&& +2682738003029262336\,{w}^{37} +297237575406452736\,{w}^{38}
\Bigr) \nonumber
\end{eqnarray}
\begin{eqnarray}
&& p_1 =\,  2\cdot \left (4\,w-1\right )\left (4\,w+1\right )\left (96\,{w}^{4}+104\,{
w}^{3}-18\,{w}^{2}-3\,w+1\right ) \nonumber \\
&& \Bigl (-3 -25\,w +1013\,{w}^{2}+7893\,{w}^{3} -353904\,{w}^{4}-1562671\,{w}^{5} \nonumber \\
&& +43285825\,{w}^{6}+192457911\,{w}^{7}  -2690351207\,{w}^{8}-15077420736\,{w}^{9} \nonumber \\
&& +94510776436\,{w}^{10} +707838800508\,{w}^{11}-2327528107216\,{w}^{12} \nonumber \\
&& -23421365465744\,{w}^{13} +45755890012000\,{w}^{14}+568028144875200\,{w}^{15} \nonumber \\
&& -824814656530816\,{w}^{16} -10390722028797440\,{w}^{17} \nonumber \\
&& +12438134957505536\,{w}^{18} +145637031330319360\,{w}^{19} \nonumber \\
&& -127616737495506944\,{w}^{20} -1708173874007113728\,{w}^{21} \nonumber \\
&& -52355400373420032\,{w}^{22} +15741676181476802560\,{w}^{23} \nonumber \\
&& +24085046332129804288\,{w}^{24} -57977682482294161408\,{w}^{25} \nonumber \\
&& -168033877030234750976\,{w}^{26} -56941336876602621952\,{w}^{27} \nonumber \\
&& -426707803148891717632\,{w}^{28} -200805832817071095808\,{w}^{29} \nonumber \\
&& +8716841486700848873472\,{w}^{30}-6642009916749838811136\,{w}^{31}       \nonumber \\
&& -192590979400145399971840\,{w}^{32} -564260086660360537374720\,{w}^{33} \nonumber \\
&& -585770764250229243904000\,{w}^{34} +235172208485444226121728\,{w}^{35} \nonumber \\
&& +1203159617695281059987456\,{w}^{36}+1323272087085206269329408\,{w}^{37} \nonumber \\
&& +997072075164663150542848\,{w}^{38}+789138181323007857786880\,{w}^{39}   \nonumber \\
&& +388137877034203055390720\,{w}^{40} +4946627729914186432512\,{w}^{41} \nonumber \\
&& -26947297377570617556992\,{w}^{42} +10614515947351012540416\,{w}^{43} \nonumber \\
&& +998718253365681192960\,{w}^{44}
\Bigr)     \nonumber
\end{eqnarray}
\begin{eqnarray}
&& p_0 =\, 
2\,w \cdot \Bigl(-348+2768\,w+248784\,{w}^{2}-358217\,{w}^{3}-50461860\,{w}^{4} \nonumber \\
&& +16394998\,{w}^{5}+5283255372\,{w}^{6}+3911764831\,{w}^{7}-329364073508\,{w}^{8} \nonumber \\
&& -572985025996\,{w}^{9}+13847002317264\,{w}^{10}+38091073842520\,{w}^{11}\nonumber \\
&& -437846238222272\,{w}^{12}-1682624909395232\,{w}^{13}            \nonumber \\
&& +10892230218721408\,{w}^{14}+52959188332189824\,{w}^{15}         \nonumber \\
&& -214291413015639808\,{w}^{16}-1200734422407578112\,{w}^{17}      \nonumber \\
&& +3319489124092462080\,{w}^{18}+20066023020568346624\,{w}^{19} \nonumber \\
&& -38248948302383529984\,{w}^{20}  -254480826931185762304\,{w}^{21} \nonumber \\
&& +261281404771497082880\,{w}^{22}+2480194764802183397376\,{w}^{23} \nonumber \\
&& +148352203759030894592\,{w}^{24} -19049822668612433870848\,{w}^{25}         \nonumber \\
&& -29328532357149024583680\,{w}^{26}+103410036785394615320576\,{w}^{27}            \nonumber \\
&& +391034390334579595542528\,{w}^{28}+11096790708133489016832\,{w}^{29}            \nonumber \\
&& -1530120948962096058466304\,{w}^{30} -2868669407093825701150720\,{w}^{31} \nonumber \\
&& -6126661019209831555268608\,{w}^{32}+2808943911875675603075072\,{w}^{33}         \nonumber \\
&& +40458568379798955017371648\,{w}^{34} -169712327643359793079386112\,{w}^{35} \nonumber \\
&&-1092943871171162347998806016\,{w}^{36} -1781375524629107822238367744\,{w}^{37} \nonumber \\
&&+250471471742289487729786880\,{w}^{38} +4679788548889591917580386304\,{w}^{39} \nonumber \\
&&+7101176295364126941625974784\,{w}^{40} +5918768536906007398653624320\,{w}^{41}\nonumber \\
&& +4083406571846803705271681024\,{w}^{42}  +2567747434748530216944009216\,{w}^{43}\nonumber \\
&&+846246487598480459424595968\,{w}^{44} -49595159800068478383161344\,{w}^{45} \nonumber \\
&&-37040268890013610134208512\,{w}^{46}  +21784239691989525951676416\,{w}^{47}   \nonumber \\
&& +1753178556765355785584640\,{w}^{48}
 \Bigr)  \nonumber 
\end{eqnarray}

\section{Appendix B: Solutions of the differential operator $Y_3$}
\label{secappa1}
Considering the critical exponents at the regular singular points, as well as
the formal solutions of differential operator $Y_3$, one can make the following
remarks. The roots of the polynomial of degree 28 in polynomial $p_3$
(see (\ref{d3})) are {\em apparent singularities}. The roots of the
polynomial of degree
four in one of the factors of the same polynomial $p_3$ are not apparent singularities.
While the formal solutions near $w=0$, $w=\pm 1/4$, and $w=\infty$, have
one Frobenius solution and two logarithmic solutions, the formal solutions
near the other regular singular points are free of logarithmic solutions.
The critical exponents at $w=1$, $w=-1/2$, roots of $1+3w+4w^2=0$, and
roots of $1-3\,w-18\,{w}^{2}+104\,{w}^{3}+96\,{w}^{4}=0$, are respectively
$(-1, 0, 1)$, $(-1, 0, 1)$, $(-1, 0, 1)$ and $(-1, 1, 2)$. This leads
us to look for the solutions of the third order differential operator $Y_3$ as
a linear combination of powers of elliptic integrals with a common factor ``taking care'' of 
 the non logarithmic singularity behavior of the singular points.

Defining
\begin{eqnarray}
K (x) = \, _{2}F_1 \left( 1/2, 1/2; 1; x \right), \quad \quad \quad
E (x) = \, _{2}F_1 \left( 1/2, -1/2; 1; x \right) \nonumber 
\end{eqnarray}
and
\begin{eqnarray}
s(w) &=& \, \, w^2 \cdot \left(1-16w^2\right)^3 
\left( 1+2\,w \right)  \left( 1-w \right)  
\left( 1+3\,w+4\,{w}^{2} \right) \nonumber \\
&& \quad 
\left(1-3\,w -18\,{w}^{2}+104\,{w}^{3}+96\,{w}^{4}\right) \nonumber 
\end{eqnarray}
one obtains the three independent solutions of
the differential operator $Y_3$ as:
\begin{eqnarray}
S_1(Y_3) &=&\, \, \, \,  {\frac{1}{s(w)}} \cdot 
\Bigl( P_{1} \cdot K^2(16w^2)\,  +P_{2}\cdot E^2(16w^2)  \nonumber \\
&& \quad \quad \quad + P_{3}\cdot  K(16w^2)\,E(16w^2)  \Bigr)
\nonumber
\end{eqnarray}
\begin{eqnarray}
S_2(Y_3) &=& \,\,\,\,{\frac{1}{s(w)}} \cdot  
\Bigl( P_{4}\cdot K^2(1/16w^2) -16w^2 P_{2} \cdot E^2(1/16w^2) \nonumber \\
&& \quad \quad +P_{5} \cdot K(1/16w^2) \, E(1/16w^2)  \Bigr)
\nonumber \\
S_3(Y_3) &=& \,\,{\frac{1}{s(w)}} \cdot 
\Bigl( (P_1+P_2+P_3)\cdot  K^2(1-16w^2) +P_{2} \cdot E^2(1-16w^2) \nonumber \\
&& \quad -(2P_{2}+P_3) \cdot K(1-16w^2)\,E(1-16w^2)  \Bigr)
\nonumber
\end{eqnarray}
with
\begin{eqnarray}
P_4 &=& -{{P_1 } \over {16 \, w^2}} \,  - {{(1-16\, w^2)^2} \over {16 \, w^2}}  \cdot P_2
- {{1-16\, w^2} \over {16 \, w^2}} \cdot P_3, \nonumber \\
P_5 &=& - 2(1-16\, w^2) \cdot P_2- P_3 \nonumber
\end{eqnarray}
 where the three polynomials $P_1$, $P_2$ and $P_3$  read:
\begin{eqnarray}
P_1 &=& 
-\left( 1+4\,w \right)
\Bigl( 1-5\,w-69\,{w}^{2}+537\,{w}^{3}+2964\,{w}^{4}-4100\,{w}^{5} \nonumber \\
&& -46816\,{w}^{6}-74688\,{w}^{7}+230656\,{w}^{8}+647680\,{w}^{9}+475136\,
{w}^{10} \nonumber \\
&& -8192\,{w}^{11}+720896\,{w}^{12} \Bigr)
    \nonumber \\
P_2 &=& 
-1+5\,w+25\,{w}^{2}-9\,{w}^{3}-2408\,{w}^{4}-17460\,{w}^{5}-19696\,{w}
^{6}  \nonumber \\
&& +28800\,{w}^{7}-3328\,{w}^{8}-62464\,{w}^{9}-36864\,{w}^{10}    \nonumber \\
P_3 &=& 
2 \cdot \Bigl( 1-3\,w-65\,{w}^{2}+143\,{w}^{3}+3888\,{w}^{4}+15144\,{w}^{
5}-10624\,{w}^{6} \nonumber \\
&& -172416\,{w}^{7}-241536\,{w}^{8}+111616\,{w}^{9}+
282624\,{w}^{10} \nonumber \\
&& +180224\,{w}^{11}+98304\,{w}^{12} \Bigr)    \nonumber 
\end{eqnarray}

{\bf Remark:} Let us note the very close similarity between the differential
operator $Y_3$, occurring at the left of differential operator $L_6$
(see (\ref{morefactoL7})) for
$\tilde{\chi}^{(3)}$, and the differential operator $\, M_2$ (see (\ref{factochi4bis}))
occurring at the left of differential operator ${\cal L}_{8}$ for
$\tilde{\chi}^{(4)}$. For this order four differential operator $M_2$, we have
been able, using the same ansatz, to obtain in closed form three of 
the four solutions, also expressed as a linear combination of products of
elliptic integrals.
Note that, setting $\lambda=16w^2$, one
can detect in the solutions of $Y_3$ (and also in the three solutions of $M_2$
we have found) the structure of $\Sigma_3$ permutation 
group \cite{Fond}, $\lambda$, $1/\lambda$,
$1-\lambda$, $\, 1-1/\lambda$, etc.

\section{Appendix C: Connection matrices between $w=0$ and $w=-1/4$, $w=\infty$}
\subsection{Connection matrix between $w=0$ and $w=-1/4$}
The basis of solutions at the anti-ferromagnetic critical point $w=-1/4$
are chosen as follows (with $x=1+4w$)
\begin{eqnarray}
S_1^{(-1/4)} (x) &=& \,  {\cal S}(N_1)(x),           \nonumber \\
S_2^{(-1/4)} (x) &=& \,  [1, 0, 1/10, -87/700, -313/1680, \cdots ],     \nonumber \\
S_3^{(-1/4)} (x) &=& \,  [0, 1, -17/10, -23/25, -1/30, \cdots ],           \nonumber \\
S_4^{(-1/4)} (x) &=& \,  [1, -5/2, -3/8, 5/16, 83/512, \cdots ],         \nonumber \\
S_5^{(-1/4)} (x) &=&\,  S_{4}^{(-1/4)}(x) \cdot \left( \ln(x/8) + 6 \right) + S_{50}^{(-1/4)}(x) \nonumber \\
S_6^{(-1/4)} (x) &=&\,  S_{4}^{(-1/4)}(x) \cdot \left( \ln^2(x/8)+12\ln(x/8) +23264/315 \right) \nonumber \\
&&    \quad     +2 S_{50}^{(-1/4)}(x) \cdot \left( \ln(x/8) +6 \right) + S_{60}^{(-1/4)}(x) \nonumber
\end{eqnarray}
with:
\begin{eqnarray}
S_{50}^{(-1/4)} (x) &=& \,  [0, 97/6, 553/240, -2339/672, -1678457/645120, \cdots ],    \nonumber \\
S_{60}^{(-1/4)} (x) &=& \, [0, 0, 0, 85997/18000, 8450503/1814400, \cdots ]. \nonumber
\end{eqnarray}
Here again, an optimal choice of the components is made in order to
remove logarithms and have as many zeroes as possible in the entries of the matrix.
The same method of matching the series-solutions at a half-way point
between $w=0$ and $w=-1/4$, gives
\begin{eqnarray}
\label{C0m14}
&& C(0,\, -1/4) \,= \\
&& \left[ \begin {array}{cccccc} 
1&0&0&0&0&0\\
\noalign{\medskip}
-2&r_{22}&r_{23}&0&0&0\\
\noalign{\medskip}
-2 \pi \, i&r_{32}+r_{22} \, \pi \, i &r_{33} +r_{23} \, \pi \, i &0&0&0\\
\noalign{\medskip}
6&{\frac {{1}}{\pi }} i_{52} &{\frac {{1}}{\pi }}i_{53} &0&0&{\frac{1}{8 \pi^2}}\\
\noalign{\medskip}
{\frac{5}{2}}+6\pi\,i 
&a_{52}
&a_{53}
&0&{\frac{1}{16}}&{\frac{1}{8 \pi}}i\\
\noalign{\medskip}
-{\frac{23}{8}}-{\frac{17\pi^2}{3}}+5\pi\,i 
&a_{62}
&a_{63}
&{\frac{\pi^2}{32}}
&{\frac{\pi}{8}}i &-{\frac{1}{8}}
\end {array} \right]
\nonumber
\end{eqnarray}
with:
\begin{eqnarray}
\,r_{22} r_{33}\, - r_{23} r_{32}\, = \, 25/12288 \nonumber \\
a_{52}= -3 r_{32} - {\frac {5}{4}} r_{22} +i_{52} i, \quad \quad 
a_{53}=-3 r_{33} - {\frac {5}{4}} r_{23} +i_{53} i \nonumber \\
a_{62}= \bigl(  {\frac{25}{16}} - {\frac{2\pi^2}{3}}
 -{\frac{5\pi}{2}} \, i  \bigr)\,  r_{22}
-\bigl(  {\frac{5}{2}} +6\pi\ i \bigr)\,  r_{32}-i_{52}\, \pi \nonumber \\
a_{63}=  \bigl(  {\frac{25}{16}} - {\frac{2\pi^2}{3}}
 -{\frac{5\pi}{2}} \, i  \bigr)\,  r_{23}
-\bigl(  {\frac{5}{2}} +6\pi\ i \bigr)\,  r_{33}\, -i_{53}\, \pi  \nonumber 
\end{eqnarray}
and where\footnote[2]{The numbers $r_{ij}$'s and $i_{ij}$'s are peculiar
to each connection matrix.}:
\begin{eqnarray}
r_{22} \simeq
-0.059050961331, \quad r_{23} \simeq -0.018643190255,  \nonumber\\
r_{32} \simeq
0.1631382423131, \quad i_{52} \simeq -1.839621665835,  \nonumber\\
i_{53} \simeq
-0.015467563102 \nonumber
\end{eqnarray}

\subsection{Connection matrix between $w=0$ and $w=\infty$}
\label{connec}
The basis of solutions at the singular point $w=\infty$
are chosen as follows (with $x=1/w$):
\begin{eqnarray}
S_1^{(\infty)} (x) &=& \,\, {\cal S}(N_1),           \nonumber \\
S_2^{(\infty)} (x) &=& \,\, [1, 1, 7/16, 1/16, 7/256, \cdots ],            \nonumber \\
S_3^{(\infty)} (x) &=& \,\,\, \left( \ln(x/4)-2/3 \right) \cdot S_2^{(\infty)}(x)\,
 +S_{30}^{(\infty)}(x), \nonumber \\
S_4^{(\infty)} (x) &=& \,\, [0, 1, 0, 1/32, -9/512, \cdots ],           \\
S_5^{(\infty)} (x) &=&\,\,  \left( \ln(x/16) +a_1 \right) \cdot S_{4}^{(\infty)}(x)
 + S_{50}^{(\infty)}(x), \nonumber \\
S_6^{(\infty)} (x) &=& \,\, \left( \ln^2(x/16)+2a_1 \ln(x/16)+a_2
 \right)\cdot S_{4}^{(\infty)}(x) \nonumber \\
&&\, +  2\, \left( \ln(x/16)+a_1 \right)\cdot S_{50}^{(\infty)}(x) +
S_{60}^{(\infty)}(x) \nonumber
\end{eqnarray}
with:
\begin{eqnarray}
a_1=-5-{{\pi}\over{2}}\,i, \qquad \quad \quad a_2=-{{\pi^2}\over{4}}+{{379}\over{11}}
+5\pi \, i, \nonumber
\end{eqnarray}
\begin{eqnarray}
S_{30}^{(\infty)} (x) &=& \,\, [2/3, 1/6, 1/24, -1/96, 7/768, \cdots ],            \nonumber \\
S_{50}^{(\infty)} (x) &=& \,\, [0, 0, -3/2, 3/64, -107/512, -23113/491520, \cdots ],            \nonumber \\
S_{60}^{(\infty)} (x) &=& \,\, [0, 0, 0, 93/44, -80891/13516, 105811/4055040, \cdots ].            \nonumber 
\end{eqnarray}
The connection matrix reads
\begin{eqnarray}
\label{C0inf}
&&C(0,\, \infty)\, =\, \,  \\
&&\left[ \begin {array}{cccccc} 
1&0&0&0&0&0\\
\noalign{\medskip}
1&-{\frac{1}{16}}&
-{\frac {3}{16 \pi }}\, i&0&0&0\\
\noalign{\medskip}
-\pi\,i &0&-{\frac{1}{16}}&0&0&0\\
\noalign{\medskip}
-11+y_{41} i
&x_{42}-{\frac {1}{\pi }} i
&{\frac{2}{\pi^2}}-{\frac {15}{16\pi}}\,i
&0&0&{\frac{1}{4 \pi^2}}\\
\noalign{\medskip}
a_{51} & a_{52}
&-{\frac {9}{16}}-{\frac {49}{64 \pi }}i
&0&{\frac{1}{16}}&-{\frac{1}{8 \pi}} i\\
\noalign{\medskip}
a_{61} &a_{62}
&-{\frac {11}{32}}+{\frac{5 \pi}{16}}i -{\frac{75}{256 \pi}} i
&{\frac {\pi^2}{64}}& -{\frac{\pi}{16}} i &-{\frac{1}{16}}
\end {array} \right]
\nonumber
\end{eqnarray}
where:
\begin{eqnarray}
&&x_{42}\, \simeq \, 
-1.534248223197, \quad \quad y_{41} \simeq -22.932479960454, \nonumber \\
&& a_{51}=-{\frac{5}{4}} +{\frac{\pi}{2}}y_{41} +7\pi \,i, \quad \quad
a_{52} =-{\frac {11}{64}}-{\frac{\pi}{2}}x_{42}i-{\frac{\pi}{32}}i, \nonumber \\
&&a_{61} \, = \, \, {\frac{29}{16}} + {\frac{16\pi^2}{3}} -{\frac{\pi^2}{4}} i\, y_{41}\, +{\frac{5 \pi}{2}}\,i,
 \quad \,  
a_{62} \, = \, \, -{\frac{25}{256}}-{\frac {7\pi^2}{192}}\,  -{\frac{\pi^2}{4}} x_{42}.  \nonumber
\end{eqnarray}

\section{Appendix D} 
\subsection{Basis of solutions for  $\, w=1$, $w=-1/2$ and $\, \, 1+3w+4w^2=0$.}
\label{appeb0}
The basis near $w=1$ is (with $x=1-w$):
\begin{eqnarray}
S_1^{(1)}(x) &=& \, \,  {\cal S}(N_1)(x),          \nonumber \\
S_2^{(1)}(x) &=& \, \,  [0, 0, 0, 1, 65/24, 383/72, \cdots],           \nonumber \\
S_3^{(1)}(x) &=& \, \, S_{2}^{(1)}(x) \cdot \left( \ln(x/24) +2666/75 \right) + S_{30}^{(1)}(x), \nonumber \\
S_4^{(1)}(x) &=&\, \,   [0, 1, 0, 0, 0, -213149176769/914630737500, \cdots],           \nonumber \\
S_5^{(1)}(x) &=&\, \,   [0, 0, 1, 0, 0, 806017240807/426827677500, \cdots],           \nonumber \\
S_6^{(1)}(x) &=&\, \,   [0, 0, 0, 0, 1, 555108887/158084325, \cdots],           \nonumber 
\end{eqnarray}
with:
\begin{eqnarray}
S_{30}^{(1)}(x) = \, \, [0, 96/5, 628/25, 0, -812657/18000,  \cdots].           \nonumber 
\end{eqnarray}

The basis near $w=-1/2$ reads (with $x=1+2w$)
\begin{eqnarray}
S_1^{(-1/2)}(x) &=& \, \,  {\cal S}(N_1)(x),          \nonumber \\
S_2^{(-1/2)}(x) &=& \, \,  [0, 0, 0, 1, 8/3, 46/9, 247/27, \cdots],           \nonumber \\
S_3^{(-1/2)}(x) &=& \, \, S_{2}^{(-1/2)}(x) \cdot \ln(x)  + S_{30}^{(-1/2)}(x), \nonumber \\
S_4^{(-1/2)}(x) &=& \, \,  [0, 1, 0, 0, 0, -55489/60345, \cdots],           \nonumber \\
S_5^{(-1/2)}(x) &=&\, \,   [0, 0, 1, 0, 0, 159977/80460, \cdots],           \nonumber \\
S_6^{(-1/2)}(x) &=& \, \,  [0, 0, 0, 0, 1, 1492/447, \cdots]           \nonumber 
\end{eqnarray}
where:
\begin{eqnarray}
S_{30}^{(-1/2)}(x) =  [0, 3/4, 7/8, 0, -95/144,  \cdots].           \nonumber 
\end{eqnarray}

The basis near $w_1=-3/8+i\,\sqrt{7}/8$ root of $1+3w+4w^2$
is (with $x=1-w/w_1$)
\begin{eqnarray}
S_1^{(w_1)}(x) &=& \, \,  {\cal S}(N_1)(x),          \nonumber \\
S_2^{(w_1)}(x) &=& \, \,  [0, 1, 49/64-61/(64\sqrt{7})\, i, 655/1024-747/(1024\sqrt{7})\, i,  \cdots], 
  \nonumber \\
S_3^{(w_1)}(x) &=& \, \, S_{2}^{(w_1)}(x) \cdot \ln(x)  + S_{30}^{(w_1)}(x), \quad 
S_4^{(w_1)}(x) = \, \,  [0, 0, 1, 0, 0,  \cdots],          \nonumber \\
S_5^{(w_1)}(x) &=& \, \,  [0, 0, 0, 1, 0,  \cdots],   \qquad 
S_6^{(w_1)}(x) = \, \,  [0, 0, 0, 0, 1,  \cdots]           \nonumber 
\end{eqnarray}
with:
\begin{eqnarray}
S_{30}^{(w_1)}(x) = \, \,
 [0, 0, 657/896+61/(128\sqrt{7})\, i, 41203/43008+1991/(6144\sqrt{7})\, i,  \cdots].  \nonumber 
\end{eqnarray}

\subsection{Connection matrices for $w=1$, $w=-1/2$ and $1+3w+4w^2=0$}
\label{appb}
For the singular point $w=1$, the connection matrix with $w=0$ reads
\begin{eqnarray}
C(0,\, 1)\, \, = \, \,
\left[ \begin {array}{cc} 
{\bf A}&{\bf 0}\\
\noalign{\medskip}
{\bf B}&{\bf C}
\end {array} \right] \quad {\rm where }\quad  \left[ \begin {array}{c} 
{\bf A}\\
\noalign{\medskip}
{\bf B}
\end {array} \right]
\quad {\rm and }\quad 
\left[ \begin {array}{c} 
{\bf C}
\end {array} \right]
\quad {\rm read}
\nonumber
\end{eqnarray}
\begin{eqnarray}
\left[ \begin {array}{ccc} 
1&0&0\\
\noalign{\medskip}
4+ i_{21} i 
&-{\frac {\sqrt {3}}{144}}\,i
&-{\frac {\sqrt {3}}{144 \pi}}\\
\noalign{\medskip}
-2 \pi \,i 
&-{\frac {\pi \sqrt {3}}{216}}\, 
&0 \\
\noalign{\medskip}
-4 -{\frac {4}{\pi }} i_{51} -{\frac {5}{\pi }} i_{21} + i_{41} i 
&r_{42}+{\frac {4}{\pi }} r_{52} i +{\frac {\sqrt {3}}{48 }} i
&-{\frac {7\sqrt {3}}{48\pi}}\\
\noalign{\medskip}
5-{\frac {2}{\pi }}i_{61}+\bigl(     {\frac {2\pi}{3}}+{\frac {25}{8\pi}}   \bigr)  \,i_{21}+i_{51}i
&r_{52}+{\frac {2}{\pi }} r_{62} i -{\frac {25 \sqrt {3}}{1728}}\,i
&{\frac {5\sqrt {3}}{576\pi}}+{\frac{\sqrt {3}}{18}} i\\
\noalign{\medskip}
{\frac {13}{2 }}+{\frac {\pi^2}{3}} +i_{61} i
&r_{62}-{\frac {\pi^2 \sqrt {3} }{432}} i -{\frac {25\sqrt {3}}{2304}} i
&{\frac {11\pi \sqrt {3}}{432}}   - {\frac {25\sqrt {3}}{2304 \pi}} 
\end {array} \right]
\nonumber
\end{eqnarray}
\begin{eqnarray}
\left[ \begin {array}{ccc} 
r_{44}+ i_{44}i
&r_{45}+ i_{45}i
&r_{46}+ i_{46}i\\
\noalign{\medskip}
{\frac {\pi}{4 }} i_{44}+i_{54}i
&{\frac {\pi}{4 }} i_{45}+i_{55}i
&{\frac {\pi}{4 }} i_{46}+i_{56}i\\
\noalign{\medskip}
{\frac {\pi}{2 }} i_{54}
&{\frac {\pi}{2 }} i_{55}
&{\frac {\pi}{2 }} i_{56}
\end {array} \right]
\nonumber
\end{eqnarray}
where:
\begin{eqnarray}
i_{21} \simeq
1.838093775180, \,\, i_{41} \simeq 4.136525226980, \,\, i_{51} \simeq -8.13898927603 \nonumber\\
i_{61} \simeq
20.74366088704, \,\, r_{42} \simeq 2.542631644752, \,\, r_{52} \simeq -0.01184208897 \nonumber\\
r_{62} \simeq
-4.87108777344, \,\, r_{44} \simeq 1.622875171987, \,\, r_{45} \simeq 1.954781507112 \nonumber\\
r_{46} \simeq
-3.51387499953, \,\, i_{44} \simeq 0.158271118920, \,\, i_{54} \simeq -2.13873967059 \nonumber\\
i_{45} \simeq
0.041310289307, \,\, i_{55} \simeq -2.46759854730, \,\, i_{46} \simeq -0.02873064396 \nonumber\\
i_{56}\simeq
4.392293882282, \,\, \nonumber
\end{eqnarray}
These numbers are such that:
\begin{eqnarray}
&&i_{46}i_{55}r_{44}\, +r_{45}i_{56}i_{44}\, -r_{46}i_{55}i_{44}\, -i_{46}
r_{45}i_{54}\nonumber \\
&& \qquad \qquad \qquad +i_{45}r_{46}i_{54}\, -i_{45}i_{56}r_{44}\, \,
=\, \, -{\frac {468398}{18984375 \pi^2}}
\nonumber
\end{eqnarray}

The connection matrix between $w=0$ and the singular point $w=-1/2$,
reads
\begin{eqnarray}
C(0,\, -1/2)\, \, = \, \,
\left[ \begin {array}{cc} 
{\bf A}&{\bf 0}\\
\noalign{\medskip}
{\bf B}&{\bf C}
\end {array} \right] \quad {\rm where }\quad  \left[ \begin {array}{c} 
{\bf A}\\
\noalign{\medskip}
{\bf B}
\end {array} \right]
\quad {\rm and }\quad 
\left[ \begin {array}{c} 
{\bf C}
\end {array} \right]
\quad {\rm read}
\nonumber
\end{eqnarray}
\begin{eqnarray}
 \left[ \begin {array}{ccc} 
1&0&0
\\
\noalign{\medskip}
-2-{\frac {3 \ln(2) }{\pi }} \, i&r_{22}&-{\frac {\sqrt {3}}{9 \pi }}
\\
\noalign{\medskip}
-\pi i+3\, \ln(2)
&-{\frac {\pi \sqrt {3} }{54}}+\pi \, r_{22} \,  i  
&{\frac {-\sqrt {3}}{9}} \, i
\\
\noalign{\medskip}
   r_{41}+i_{41}i
&r_{42}+i_{42}i
&{\frac {11 \sqrt {3} }{9 \pi}}
\\
\noalign{\medskip}
   r_{51}+i_{51}i
&r_{52}+i_{52}i
&{\frac {5 \sqrt {3} }{36 \pi }}+{\frac {7\sqrt {3}}{9}} i
\\
\noalign{\medskip}
   r_{61}+i_{61}i
&r_{62}+i_{62}i
&-{\frac {25 \sqrt {3}}{144\pi}} -{\frac {13 \pi \sqrt {3}}{27}} +{\frac {5\sqrt {3}}{18}} i   
\end {array} \right],
\nonumber
\end{eqnarray}
\begin{eqnarray}
 \left[ \begin {array}{ccc} 
r_{44}+i_{44}i
&r_{45}+i_{45}i
&r_{46}+i_{46}i\\
\noalign{\medskip}
-{\frac {3 \pi }{4}} i_{44}+i_{54}i
&-{\frac {3 \pi }{4}}i_{45}+i_{55}i
&-{\frac {3 \pi }{4}}i_{46}+i_{56}i\\
\noalign{\medskip}
r_{64}-{\frac { \pi^2 }{2}} i_{44}i   
&r_{65}-{\frac { \pi^2 }{2}} i_{45}i
&r_{66}-{\frac { \pi^2 }{2}} i_{46}i
\end {array} \right]
\nonumber
\nonumber
\end{eqnarray}
where:
\begin{eqnarray}
r_{22}\simeq
-0.02539959775, \,\, r_{41}\simeq 6.805351589429, \,\, r_{51} \simeq 7.203810787172, \nonumber\\
r_{61}\simeq
-8.75798651623, \,\, i_{41}\simeq -5.23529215352, \,\, i_{51} \simeq 12.14972643902, \nonumber\\
i_{61}\simeq
7.505979318469, \,\, r_{42}\simeq 0.512271205543, \,\, r_{52} \simeq -0.75497554989, \nonumber\\
r_{62}=
2.232400538972, \,\, i_{42}\simeq 0.462196540081, \,\, i_{52} \simeq 0.143220115658, \nonumber\\
i_{62}\simeq
-0.18195427623, \,\, r_{44}\simeq -0.1681290553, \,\, r_{45} \simeq -0.00270658055, \nonumber\\
r_{46}\simeq
-0.00323043290, \,\, i_{44}\simeq -0.14301292413, \,\, i_{45}\simeq 0.690508507395, \nonumber\\
i_{46}\simeq
-1.26354926677, \,\, i_{54}\simeq -0.34844554701, \,\, r_{64}\simeq 0.812327323812, \nonumber\\
i_{55}\simeq
-0.50108648504, \,\, r_{65}\simeq 2.347957990666, \,\, i_{56}\simeq 1.132041888142, \nonumber\\
r_{66}\simeq
-5.35056326640, \,\, \nonumber
\end{eqnarray}

The connection matrix between $w=0$ and the singular point
$w_1=-3/8 +i\,\sqrt{7}/8$ root of $\, 1+3w+4w^2=0$, reads
\begin{eqnarray}
C(0,\, w_1)= 
\left[ \begin {array}{cc} 
{\bf A}&{\bf 0}\\
\noalign{\medskip}
{\bf B}&{\bf C}
\end {array} \right] \quad \hbox{where} \quad 
\left[ \begin {array}{c} 
{\bf A}\\
\noalign{\medskip}
{\bf B}
\end {array} \right]
\quad \hbox{and} \quad \left[ \begin {array}{c} 
{\bf C}
\end {array} \right] \quad \hbox{read}  
\nonumber
\end{eqnarray}
\begin{eqnarray}
 \left[ \begin {array}{ccc} 
1&0&0\\
\noalign{\medskip}
r_{21}-{\frac {3}{2\pi }} r_{31} i
&r_{22}-{\frac {3}{2\pi }} r_{32} i +{\frac {275\sqrt {7}}{16384}} i
&a
\\
\noalign{\medskip}
r_{31} + {\frac{2\pi}{3}} r_{21} i +{\frac{2 \pi}{3}} i
& r_{32}+{\frac{2 \pi}{3}} r_{22} i-{\frac {623 \pi }{24576}} i
& {\frac{2 \pi}{3}}i \, a
\\
\noalign{\medskip}
r_{41}+i_{41}i
&r_{42}+i_{42}i
&-{\frac{1}{3}} a
\\
\noalign{\medskip}
r_{51}+i_{51}i
&r_{52}+i_{52}i
&  ( -{\frac{5}{4}} + {\frac{2\pi}{3}}i ) a
\\
\noalign{\medskip}
r_{61}+i_{61}i
&r_{62}+i_{62}i
&   ( {\frac{25}{16}} -\pi^2 - {\frac{5\pi}{3}}i )  a
\end {array} \right]
\nonumber
\end{eqnarray}
\begin{eqnarray}
 \left[ \begin {array}{ccc} 
r_{44}+i_{44}i
&r_{45}+i_{45}i
&r_{46}+i_{46}i
\\
\noalign{\medskip}
r_{54}+i_{54}i
&r_{55}+i_{55}i
&r_{56}+i_{56}i
\\
\noalign{\medskip}
r_{64}+i_{64}i
&r_{65}+i_{65}i
&r_{66}+i_{66}i
\end {array} \right]
\nonumber
\end{eqnarray}
where:
\begin{eqnarray}
a= {\frac {825\sqrt {7}-1869 i}{16384 \pi}}, \nonumber\\
r_{21} \simeq 
-0.30983963151, \,\, r_{31} \simeq 1.38629436111, \,\, r_{22} \simeq -0.07996746793, \nonumber\\
r_{32} \simeq 
0.044743829620, \,\, r_{41} \simeq 4.70316610599, \,\, i_{41} \simeq -5.10203220992, \nonumber\\
r_{42} \simeq 
0.028522637766, \,\, i_{42} \simeq 0.03731267544, \,\, r_{51} \simeq 1.404170417754, \nonumber\\
i_{51} \simeq 
10.77185269595, \,\, r_{52} \simeq 0.25654299002, \,\, i_{52} \simeq -0.03695328252, \nonumber\\
r_{61} \simeq 
-6.98898250954, \,\, i_{61} \simeq -17.585497074, \,\, r_{62} \simeq -0.18342705750, \nonumber\\
i_{62} \simeq 
1.339914984659, \,\, r_{44} \simeq 0.00394832042, \,\, i_{44} \simeq 0.043931830095, \nonumber\\
r_{45} \simeq 
-0.02716280332, \,\, i_{45} \simeq -0.0900753899, \,\, r_{46} \simeq 0.070134204478, \nonumber\\
i_{46} \simeq 
0.050869745772, \,\, r_{54} \simeq -0.2122947699, \,\, i_{54} \simeq 0.033562029788, \nonumber\\
r_{55} \simeq 
0.496361798471, \,\, i_{55} \simeq 0.00455966493, \,\, r_{56} \simeq -0.36867647137, \nonumber\\
i_{56} \simeq 
0.040697038977, \,\, r_{64} \simeq -0.1279407612, \,\, i_{64} \simeq -0.68382860060, \nonumber\\
r_{65} \simeq 
-0.14739127007, \,\, i_{65} \simeq 1.64596123266, \,\, r_{66} \simeq 0.189914623980, \nonumber\\
i_{66} \simeq 
-1.29483325656, \,\, \nonumber
\end{eqnarray}

\section{Appendix E: Monodromy matrices in the $\, w=0$-basis}
\label{appzz}
The monodromy matrix around $\, w=0$
expressed in terms of its own $\, (w=0)$ well-suited basis is
given in (\ref{localmono}).

The monodromy matrix around 
 $\, w\, = \, -1/2$,  expressed 
in terms of the $\, (w=0)$ well-suited basis, after a conjugation 
similar to (\ref{de0a1}), and thus using the previously 
given connection matrices, reads in terms of $\, \alpha$
and  $\, \Omega$:
\begin{eqnarray}
4 \alpha^2  \cdot  M_{w=0}(-1/2)(\alpha, \, \Omega)\,\, = \, \, 
\left[ \begin {array}{cc} 
{\bf A}&{\bf 0}\\\noalign{\medskip}
{\bf B}&{\bf C}
\end {array} \right]
\nonumber
\end{eqnarray}
where:
\begin{eqnarray}
\left[ \begin {array}{c} 
{\bf A}\\
\noalign{\medskip}
{\bf B}
\end {array} \right] =
\left[ \begin {array}{ccc} 
4\,{\alpha}^{2}&0&0\\
\noalign{\medskip}
48\,\alpha\,\Omega&4\,\alpha\, \left( 12\,\Omega+\alpha \right) &-96\,\Omega
\\
\noalign{\medskip}
24\,\Omega\,{\alpha}^{2}&24\,\Omega\,{\alpha}^{2}&4 \, \left(\alpha-12\,\Omega \right)\,\alpha \\
\noalign{\medskip}
-528\,\alpha\,\Omega&-528\,\alpha\,\Omega&1056\,\Omega\\
\noalign{\medskip}
-12\, \left( 14\,\alpha+5 \right)\,\alpha\,\Omega\ &
-12\,  \left( 14\,\alpha+5 \right)\, \alpha\,\Omega
 &24 \, \left( 14\,\alpha+5 \right)\,\Omega \\
\noalign{\medskip}
-\alpha\, a\,  \Omega&-\alpha\, a \, \Omega&2\,\Omega\, a 
\end {array} \right]
\nonumber
\end{eqnarray}
with $\, a\,=\,\left( -75+52\,{\alpha}^{2}+60\,\alpha \right)$ and
$\,
\left[ \begin {array}{c} 
{\bf C}
\end {array} \right]\,= \,\,4\,{\alpha}^{2}\cdot  {\bf Id(3\times3)} 
$.

The monodromy matrix around 
 $\, w\, = \, 1/4$,  expressed 
in terms of the $\, (w=0)$-well suited basis reads:
\begin{eqnarray}
24 \alpha^4 \cdot M_{w=0}(1/4)(\alpha, \, \Omega)\,\, = \, \,  
\left[ \begin {array}{cc} 
{\bf A}&{\bf 0}\\
\noalign{\medskip}
{\bf B}&{\bf C}
\end {array} \right]
\nonumber
\end{eqnarray}
where
$
\left[ \begin {array}{c} 
{\bf A}\\
\noalign{\medskip}
{\bf B}
\end {array} \right] 
$ 
and
$
\left[ \begin {array}{c} 
{\bf C}
\end {array} \right]
$ read respectively:
\begin{eqnarray}
 \left[ \begin {array}{ccc} 
-24\,{\alpha}^{4}&0&0
\\
\noalign{\medskip}
-48\,{\alpha}^{4}&24\,{\alpha}^{4}&-144\,{\alpha}^{2}\Omega
\\
\noalign{\medskip}
0&0&24\,{\alpha}^{4}
\\
\noalign{\medskip}
-48\, \left( 5\,{\alpha}^{4}+8\,{\Omega}^{2}+8\,{\Omega}^{2}{\alpha}^{2} \right) 
&32\, \left(4\,\Omega\,{\alpha}^{2} -75\,\Omega-15\,{\alpha}^{2} \right) \,\Omega
&48 \, \left( 9\,{\alpha}^{2}+80\,\Omega \right) \,\Omega
\\
\noalign{\medskip}
12 \,  \left( 5\,{\alpha}^{2}+4\,\Omega+4\,\Omega\,{\alpha}^{2} \right)\,{\alpha}^{2} 
&4 \, \left( 75-4\,{\alpha}^{2} \right) \,{\alpha}^{2}\Omega
&-300\,{\alpha}^{2}\Omega
\\
\noalign{\medskip}
- \left( 87+8\,{\alpha}^{2} \right)\,  {\alpha}^{4} 
&0&3 \, \left(4\,{\alpha}^{2} -75 \right)\,{\alpha}^{2}\Omega
\end {array}
 \right],
\nonumber
\end{eqnarray}
and:
\begin{eqnarray}
 \left[ \begin {array}{ccc} 
24\,{\alpha}^{4}&-384\,{\alpha}^{2}\Omega&1536\,{\Omega}^{2}
\\
\noalign{\medskip}
0&24\,{\alpha}^{4}
&-192\,{\alpha}^{2}\Omega
\\
\noalign{\medskip}
0&0&24\,{\alpha}^{4}
\end {array}
 \right]
\nonumber
\end{eqnarray}

The monodromy matrix around 
 $\, w\, = \, -1/4$,  expressed 
in terms of the $\, (w=0)$ well-suited basis reads:
\begin{eqnarray}
12\alpha^4 \cdot M_{w=0}(-1/4)(\alpha, \, \Omega)\, = \, \, 
\left[ \begin {array}{cc} 
{\bf A}&{\bf 0}\\\noalign{\medskip}
{\bf B}&{\bf C}
\end {array} \right]
\nonumber
\end{eqnarray}
where
$
\left[ \begin {array}{c} 
{\bf A}\\
\noalign{\medskip}
{\bf B}
\end {array} \right] 
$ 
and
$
\left[ \begin {array}{c} 
{\bf C}
\end {array} \right]
$ read respectively:
\begin{eqnarray}
 \left[ \begin {array}{ccc} 
-12\,{\alpha}^{4}&0&0
\\
\noalign{\medskip}
48\,{\alpha}^{4}&12\,{\alpha}^{4}&0
\\
\noalign{\medskip}
24\,{\alpha}^{5}&0&12\,{\alpha}^{4}
\\
\noalign{\medskip}
a_{41} &a_{42} &192\,\Omega\, \left( 10\,\Omega-3\,{\alpha}^{2} \right)
\\
\noalign{\medskip}
a_{51} &a_{52} &48\,\alpha\,\Omega\, \left( -5\,\alpha+20\,\Omega-6\,{\alpha}^{2} \right)
\\
\noalign{\medskip}
a_{61} &a_{62} &48\,{\alpha}^{2}\Omega\, \left( -5\,\alpha+10\,\Omega-3\,{\alpha}^{2} \right)
\end {array} \right],
\nonumber
\end{eqnarray}
with:
\begin{eqnarray}
&&a_{41}=-144\,{\alpha}^{4}-192\,{\Omega}^{2}-192\,{\Omega}^{2}{\alpha}^{2},
\nonumber \\
&&a_{42}=-16\,\Omega\, \left( 60\,\alpha\,\Omega+75\,\Omega
+8\,\Omega\,{\alpha}^{2}-18\,{\alpha}^{3}+15\,{\alpha}^{2} \right), 
\nonumber \\
&&a_{51}=-12\,\alpha\, \left( 5\,{\alpha}^{3}+6\,{\alpha}^{4}
+8\,{\Omega}^{2}+8\,{\Omega}^{2}{\alpha}^{2}-2\,\alpha\,\Omega-2\,{\alpha}^{3}\Omega \right), 
\nonumber \\
&&a_{52}=-2\,\alpha\,\Omega\, \left( 300\,\Omega+32\,\Omega\,{\alpha}^{2}
+240\,\alpha\,\Omega-80\,{\alpha}^{3}-75\,\alpha \right), 
\nonumber \\
&&a_{61}=-{\alpha}^{2} \left( -69\,{\alpha}^{2}+60\,{\alpha}^{3}
+34\,{\alpha}^{4}+48\,{\Omega}^{2}+48\,{\Omega}^{2}{\alpha}^{2}
-24\,\alpha\,\Omega-24\,{\alpha}^{3}\Omega \right), 
\nonumber \\
&&a_{62}=-2\,{\alpha}^{2}\Omega\, \left( 150\,\Omega-30\,{\alpha}^{2}
+16\,\Omega\,{\alpha}^{2}+120\,\alpha\,\Omega-44\,{\alpha}^{3}-75\,\alpha \right), 
\nonumber
\end{eqnarray}
and:
\begin{eqnarray}
 \left[ \begin {array}{ccc} 
12\, \left( \alpha+4\,\Omega \right)^{2}\,{\alpha}^{2} 
&-192 \, \left( \alpha+4\,\Omega \right)\,\alpha\,\Omega 
&768\,{\Omega}^{2}
\\\noalign{\medskip}
24\,{\alpha}^{3}\Omega\, \left( \alpha+4\,\Omega \right) 
&-12\,{\alpha}^{2} \left( -{\alpha}^{2}+32\,{\Omega}^{2} \right) 
&96\, \left( -\alpha+4\,\Omega \right) \alpha\,\Omega
\\\noalign{\medskip}
48\,{\alpha}^{4}{\Omega}^{2}
&48 \, \left(\alpha-4\,\Omega \right) \,{\alpha}^{3}\Omega
&12\, \left( -\alpha+4\,\Omega \right) ^{2}{\alpha}^{2}
\end {array} \right]
\nonumber
\end{eqnarray}

The monodromy matrix around 
 $\, w\, = \, \infty$,  expressed 
in terms of the $\, w=0$-well suited basis reads:
\begin{eqnarray}
24 \alpha^4 \cdot M_{w=0}(\infty)(\alpha, \, \Omega)\, = \, \,  
\left[ \begin {array}{cc} 
{\bf A}&{\bf 0}\\\noalign{\medskip}
{\bf B}&{\bf C}
\end {array} \right]
\nonumber
\end{eqnarray}
where
$
\left[ \begin {array}{c} 
{\bf A}\\\noalign{\medskip}
{\bf B}
\end {array} \right] 
$ 
and
$
\left[ \begin {array}{c} 
{\bf C}
\end {array} \right]
$ read respectively:
\begin{eqnarray}
 \left[ \begin {array}{ccc} 
24\,{\alpha}^{4}&0&0
\\\noalign{\medskip}
-288\,{\alpha}^{3}\Omega
&-24\,{\alpha}^{3} \left( -\alpha+6\,\Omega \right) 
&-864\,\Omega\,{\alpha}^{2}
\\\noalign{\medskip}
48\,{\alpha}^{4}\Omega
&24\,{\alpha}^{4}\Omega
&24\,{\alpha}^{3} \left( \alpha+6\,\Omega \right) 
\\
\noalign{\medskip}
a_{41} &a_{42}&96\, \left( -21\,{\alpha}^{2}+160\,\Omega \right)\,\Omega
\\
\noalign{\medskip}
a_{51} &a_{52} &-120\, \left( -6\,{\alpha}^{2}-\alpha+32\,\Omega \right)\,\alpha\,\Omega 
\\
\noalign{\medskip}
a_{61} &a_{62} &6\, \left( 20\,\alpha-225-36\,{\alpha}^{2}+160\,\Omega \right) \,{\alpha}^{2}\Omega
\end {array} \right]
\nonumber
\end{eqnarray}
with:
\begin{eqnarray}
&&a_{41}=96\, \left( {\alpha}^{3}+16\,\Omega\,{\alpha}^{2}-16\,\Omega \right) \,\Omega\,
\nonumber \\
&&a_{42}=16\, \left( -33\,{\alpha}^{3}-60\,{\alpha}^{2}
+240\,\alpha\,\Omega+8\,\Omega\,{\alpha}^{2}-600\,\Omega \right)\,\Omega, 
\nonumber \\
&&a_{51}=-24 \, \left( 2\,{\alpha}^{3}-15\,{\alpha}^{2}
-4\,\alpha+16\,\Omega\,{\alpha}^{2}-16\,\Omega \right)\,\alpha\,\Omega, 
\nonumber \\
&&a_{52}=-4 \, \left( -40\,{\alpha}^{3}-45\,{\alpha}^{2}
-150\,\alpha+240\,\alpha\,\Omega+8\,\Omega\,{\alpha}^{2}-600\,\Omega \right)\,\alpha\,\Omega,  
\nonumber \\
&&a_{61}=6 \, \left( -20\,{\alpha}^{2}-83\,\alpha
+4\,{\alpha}^{3}+16\,\Omega\,{\alpha}^{2}-16\,\Omega \right)\,{\alpha}^{2}\Omega, 
\nonumber \\
&&a_{62}= \, \left( -525\,\alpha-44\,{\alpha}^{3}
+240\,\alpha\,\Omega+8\,\Omega\,{\alpha}^{2}-600\,\Omega \right) {\alpha}^{2}\Omega,
\nonumber
\end{eqnarray}
and:
\begin{eqnarray}
 \left[ \begin {array}{ccc} 
24\, \left( -\alpha+4\,\Omega \right) ^{2}{\alpha}^{2}
&768\,\Omega\, \left( -\alpha+4\,\Omega \right) \alpha
&6144\,{\Omega}^{2}
\\\noalign{\medskip}
-24\,{\alpha}^{3}\Omega\, \left( -\alpha+4\,\Omega \right) 
&-24\,{\alpha}^{2} \left( -{\alpha}^{2}+32\,{\Omega}^{2} \right)
&-384\,\alpha\,\Omega\, \left( \alpha+4\,
\Omega \right) 
\\\noalign{\medskip}
24\,{\alpha}^{4}{\Omega}^{2}
&48\,{\alpha}^{3}\Omega\, \left( \alpha+4\,\Omega \right) 
&24\,{\alpha}^{2} \left( \alpha+4\,\Omega \right) ^{2}
\end {array} \right]
\nonumber
\end{eqnarray}


\vskip 0.5cm

\vskip 0.5cm

\begin{thebibliography}{99}


\bibitem{wu-mc-tr-ba-76}  T.T. Wu, B.M. McCoy, C.A. Tracy and E. Barouch,
1976 Phys. Rev. {\bf B 13}, 316-374 

\bibitem{nappi-78} C. R. Nappi, 1978  Nuovo Cim. A \textbf{44} 392

\bibitem{pal-tra-81} J. Palmer, C. Tracy, 1981,
 Adv. Appl. Math. \textbf{2} 329

\bibitem{yamada-84} K. Yamada, 1984 Prog. Theor. Phys. \textbf{71} 1416

\bibitem{yamada-85} K. Yamada, 1985 Phys. Lett. A \textbf{112} 456-458

\bibitem{nickel-99} B. Nickel, 1999,
 J. Phys. A: Math. Gen. \textbf{32} 3889-3906

\bibitem{nickel-00} B. Nickel, 2000,
 J. Phys. A: Math. Gen. \textbf{33}  1693-1711

\bibitem{ze-bo-ha-ma-04} N. Zenine, S. Boukraa, S. Hassani, J.M. Maillard,
2004  J. Phys. A: Math. Gen. {\bf 37} 9651-9668 and  arXiv:math-ph/0407060

\bibitem{ze-bo-ha-ma-05} N. Zenine, S. Boukraa, S. Hassani, J.M. Maillard,
{\em Square lattice Ising model susceptibility: Series expansion method
and differential equation for $\chi^{(3)}$}, 2005 
J. Phys. A: Math. Gen. {\bf 38} 1875-1899
 and arXiv:hep-ph/0411051

\bibitem{ze-bo-ha-ma-05b} N. Zenine, S. Boukraa, S. Hassani, J.M. Maillard,
{\em Ising model susceptibility: Fuchsian differential 
          equation for $\chi^{(4)}$ and its factorization properties},
          2005, J. Phys. A: Math. Gen. {\bf 38} 4149-4173
 and  cond-mat/0502155  


\bibitem{Moscou} V.I. Arnold, M. Monarstyrsky,  
 {\em Hilbert's twenty-first 
problem for Fuchsian linear systems}
in {\em Developments in Mathematics. The Moscow School.}
 V.I. Arnold and M. Monarstyrsky, editors,
(Chapman and Hall, 1993)


\bibitem{SingUlm} O. Cornier, M. F. Singer, F.  Ulmer
{\em Computing the Galois group of a polynomial
using linear differential equations},
Proceedings ISSAC 2000, 7-85


\bibitem{Stokes} B.L.J. Braaksma, G.K. Immink and M. van der Put,
{\em The Stokes phenomenon and Hilbert's 16th Problem}, (1995)
World Scientific


\bibitem{Weil} J-A Weil, N. Zenine, S. Boukraa, S. Hassani, J.M. Maillard, 
{\em Differential Galois groups associated
 with the Ising model susceptibility}, in preparation

\bibitem{or-ni-gu-pe-01b} W.P. Orrick, B.G. Nickel, A.J. Guttmann, J.H.H.
Perk, 2001 J. Stat. Phys. \textbf{102}  795-841

\bibitem{or-ni-gu-pe-01} W.P. Orrick, B.G. Nickel, A.J. Guttmann, J.H.H.
Perk, 2001 Phys. Rev. Lett. \textbf{86} 4120-4123


\bibitem{wim-zei-85}
J. Wimp and D. Zeilberger,
J. Math. Anal. App. {\bf 111} (1985) 162

\bibitem{BermSing} P. Berman, M. Singer,
 {\em Calculating the Galois Group of $\, L_1(L_2(y)) = 0$,
$\, L_1$, $\, L_2$ Completly Reducible Operators},
J. Pure Appl. Alg. {\bf 139} (1999)3-24; \\
http://www4.ncsu.edu/~singer/papers/12Inhom.ps

\bibitem{Berm} P. Berman,
{\em Computing Galois groups for certain classes of ordinary
differential equations}, Thesis, North-Carolina State University, 2001


\bibitem{Weil1}  F. Ulmer, J.-A. Weil, 
{\em Note on Kovacic's algorithm},  J. Symbolic Comput.  22  (1996),  no. 2, 179;
 J-A. Weil, {\em Absolute Factorization of
 Differential Operators},\\
http://pauillac.inria.fr/algo/seminars/sem96-97/weil.html


\bibitem{Weil2} M. van Hoeij, J-F Ragot,  F. Ulmer, J.-A. Weil, 
{\em  Liouvillian solutions of linear differential equations of order three and higher. 
Differential algebra and differential equations}.  J. Symbolic Comput.  28  (1999),  no. 4-5, 589--609 

\bibitem{Weil3}  D. Boucher,  J.-A. Weil, 
{\em  Application of J.-J. Morales and J.-P. Ramis' theorem to test the 
non-complete integrability of the planar three-body problem}.  From combinatorics 
to dynamical systems,  163--177, IRMA Lect. Math. Theor. Phys., 3, de Gruyter, Berlin, 2003 


\bibitem{Alexa} A. van der Waal, {\em Lam\'e Equations with finite monodromy}, 
Thesis Universiteit Utrecht (2002),\\
http://www.library.uu.nl/digiarchief/dip/diss/2002-0530-113355/full.pdf



\bibitem{gfun} Mgfun's project : see http://algo.inria.fr/chyzak/mgfun.html; 
gfun - generating functions package see
 gfun in : http://algo.inria.fr/libraries. 


\bibitem{salvy-91}
B. Salvy,
{\em Asymptotique automatique et fonctions g\'en\'eratrices},
PhD, Ecole Polytechnique (1991)

\bibitem{salvy-91b}
B. Salvy, {\em Examples of automatic asymptotic expansions}, SIGSAM Bulletin, Vol. 25, no 2, April (1991)


\bibitem{co-tr-wu-77}
B.M. McCoy, C.A. Tracy and T.T. Wu,
Phys. Rev. Lett. {\bf 38} (1977) 793 

\bibitem{abraham-77}
A.B. Abraham,
Phys. Lett. A {\bf 61} (1977) 271


\bibitem{Tracy} C.A. Tracy, {\em Painlev\'e transcendents and scaling functions
of the two-dimensional Ising model}, Non-linear Equations in Physics and Mathematics, 
ed. A.O. Barut, pp. 221-237, D. Reidel Publishing Co.Dordrecht, Holland, 1978 

\bibitem{Tanguy}  C. Krattenthaler, T. Rivoal, {\em Hypergeometrie et fonction zeta de Riemann}, 
 submitted to Memoirs of the AMS, arXiv.org:math.NT/0311114 


\bibitem{Tanguy2}  C. Krattenthaler, T. Rivoal, {\em An identity of Andrews,
 multiple integrals, and very-well-poised hypergeometric series }, submitted to Ramanujan Journal,
arXiv.org:math.CA/0312148

\bibitem{Tanguy3}  C. Krattenthaler, T. Rivoal, W. Zudilin,
 {\em Series hypergeometriques basiques, fonction q-zeta et series d'Eisenstein}, 
to be published in Journal de l'Institut de Mathematiques de Jussieu (2005), arXiv.org:math.NT/0311033

\bibitem{Tanguy4}  T. Rivoal, 
 {\em Series hypergeometriques et irrationalite des valeurs de la fonction zeta de Riemann}, 
 J. Theorie des Nombres de Bordeaux, 15.1 (2003), 351-365


\bibitem{Fond} J-M. Maillard and S. Boukraa, {\em Modular 
invariance in lattice statistical mechanics}, (2001) Annales de La Fondation 
Louis de Broglie, {\bf 26}, 287-328, Num\'ero Sp\'ecial



\end{thebibliography}
\end{document}